\documentclass[draftcls,onecolumn,peerreview]{IEEEtran}
\usepackage{subfigure}
\usepackage{times,amsmath,epsfig}
\usepackage{latexsym,amssymb}
\usepackage{graphicx,amssymb,amstext,amsmath}
\usepackage{setspace}
\usepackage{amsmath}
\usepackage{amsthm}
\usepackage{eucal}
\usepackage{amssymb}
\usepackage{mathrsfs}
\usepackage{color}
\usepackage{url}
\usepackage{hyperref}
\newtheorem{theorem}{Theorem}
\newtheorem{definition}{Definition}
\newtheorem{corollary}{Corollary}
\newtheorem{lemma}{Lemma}

\def\Pr{{\mathbb P}}

\def\qed{\hfill{$\Box$} \\}

\def\12{\frac{1}{2}}

\newfont{\bbb}{msbm10 scaled 500}

\newfont{\bb}{msbm10 scaled 1100}

\newcommand{\Oc}{{\cal O}}

\newtheorem{remark}{Remark}

\begin{document}

\sloppy

\title{On the Secrecy Capacity of Block Fading Channels with a Hybrid Adversary} 

\author{
  \IEEEauthorblockN{Y.~Ozan Basciftci,}
  \and
  \IEEEauthorblockN{Onur Gungor,}
  \and
  \IEEEauthorblockN{C.~Emre Koksal,}
  \and
  \IEEEauthorblockN{Fusun Ozguner}
\thanks{The authors are with the Department of Electrical and Computer Engineering, The Ohio State University, Columbus, OH 43210, USA. Email: {bascifty, gungoro, koksal,
  ozguner}@ece.osu.edu}
\thanks{This work was presented in part in the IEEE International Symposium on Information Theory (ISIT), Istanbul, July, 2013~\cite{conf_version}.}
\thanks{This work is supported in part by QNRF under grant NPRP 5-559-2-227,
and by NSF under grants CNS-1054738, CNS-0831919, CCF-0916664, and
ECCS-0931669.}
}



\maketitle
\begin{abstract}
We consider a block fading wiretap channel, where a 
transmitter attempts to send messages securely to a receiver in the presence of
a hybrid half-duplex adversary, which arbitrarily decides to either jam or eavesdrop the transmitter-to-receiver channel.
We provide  bounds to the secrecy capacity for various possibilities
on receiver feedback and show special cases where the bounds are tight.
We show that, without any feedback from the receiver, the secrecy capacity is zero
if the transmitter-to-adversary channel stochastically dominates the \emph{effective} transmitter-to-receiver channel. However, the secrecy capacity is non-zero 
even when the receiver is allowed to feed back only one bit at the end of each block. Our novel achievable strategy improves the rates proposed in the literature for the non-hybrid adversarial model. We also analyze the effect of multiple adversaries and delay constraints on the secrecy capacity. We show that our novel   time sharing approach leads to positive secrecy rates even under strict delay constraints.
\end{abstract}
\section{Introduction}
\begin{figure}[t]
   \centering
   \includegraphics[width=0.50\textwidth]{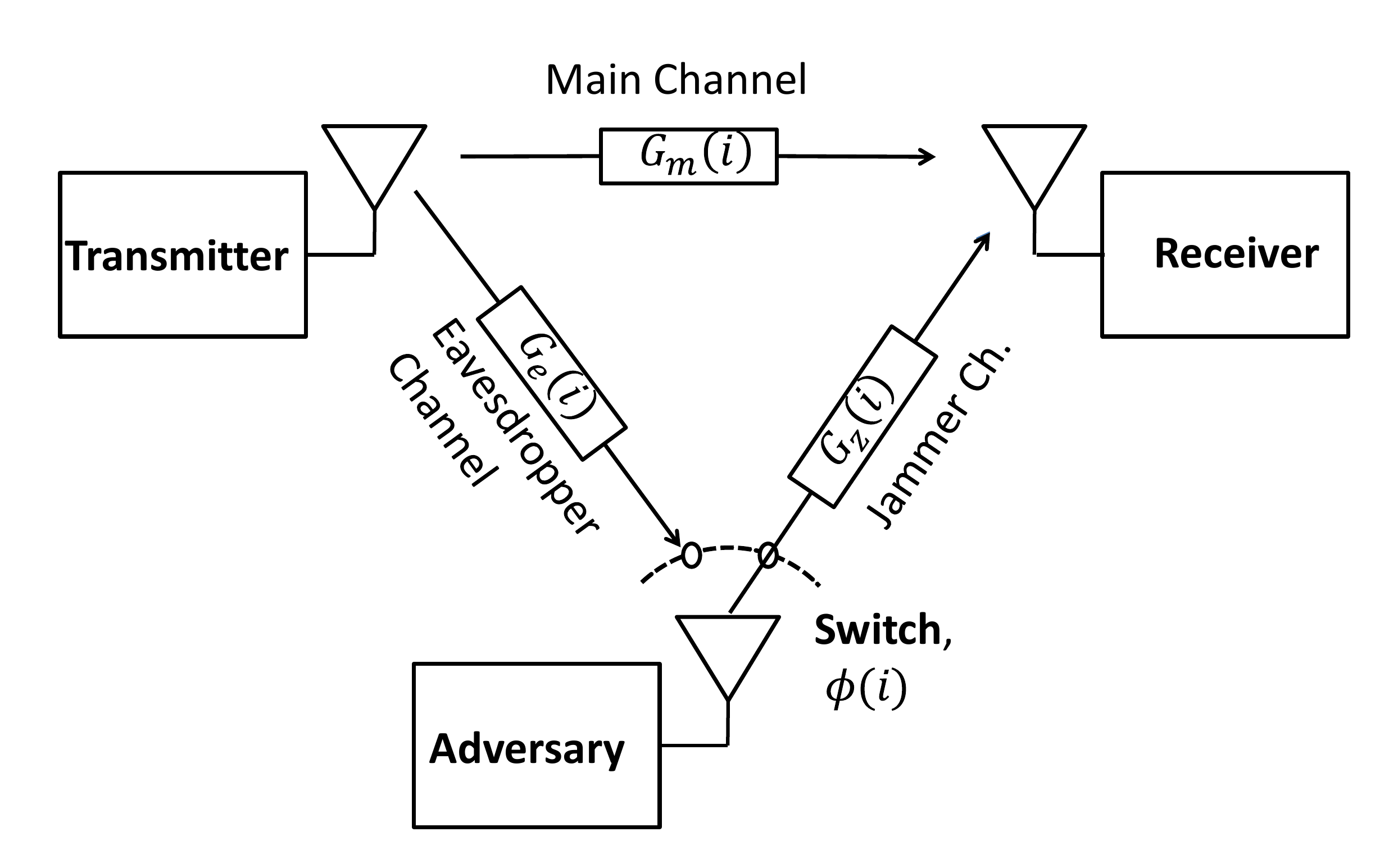}
   \caption{System Model}
   \label{fig:sysmdl}
 \end{figure}

We study point-to-point block fading channels, depicted in Figure~\ref{fig:sysmdl}, in the presence of a hybrid adversary. 
The hybrid half-duplex adversary can choose to either eavesdrop or jam the transmitter-receiver channel, but not both at a given block. The goal of the transmitter is to communicate a message reliably to the receiver while keeping it asymptotically secret from the hybrid adversary.
During the communication, the state of the adversary (jamming or eavesdropping) changes in an \emph{arbitrary} manner from one block to the next and is \emph{unknown} to the transmitter.
We further assume that the transmitter has \emph{no channel state information} (CSI) of the transmitter-to-receiver channel (main channel), the transmitter-to-adversary channel (eavesdropper channel) and the adversary-to-receiver channel (jamming channel). The receiver has perfect causal CSI of the main and jamming channels. We study the secrecy capacity of this setting when (i) there is no receiver-to-transmitter feedback, and (ii) there is $1$-bit of receiver-to-transmitter feedback sent at the \emph{end of each block}.

The main challenge in our problem stems from the fact that simultaneously maintaining reliability and secrecy is difficult because of the adversary's arbitrary strategy in choosing its state, i.e., jamming or eavesdropping, at each block. If we design a scheme focusing on a particular adversary strategy, with a slight change in that particular strategy, the adversary can cause a decoding error or a secrecy leakage. For instance, if our scheme assumes a fully eavesdropping adversary, then jamming even in a small fraction of the time will lead to a decoding error. Likewise, if the scheme is designed against a full jammer, then the adversary will lead to a secrecy leakage even it eavesdrops for a small fraction of time. A robust scheme should take into account the entire set of adversary strategies to maintain reliability and secrecy.
Our technical contributions are summarized as follows:
\begin{itemize}
\item 
We show 
that the secrecy capacity is zero when the receiver feedback is not available and
 the eavesdropper channel
stochastically dominates the \emph{effective} main channel gain. However, we also show that even one bit of receiver feedback at the end of each block
is sufficient to make the secrecy capacity positive for almost all possible channel distributions.
\item 
Under an arbitrary adversarial strategy, the receiver cannot employ a well known typical set decoder \cite{cover1991} since it cannot assume a certain distribution for the received signal. To that end, we propose a receiver strategy in which the receiver generates artificial noise and adds it to the received signal (i.e., jams itself to involve typical set decoding \cite{cover1991}). We show special cases in which artificial noise generation at the receiver is an optimal way to achieve the secrecy capacity.
\item For the 1-bit receiver feedback case, we propose a proof technique for the equivocation analysis, that is based on renewal theory. By this technique, we can improve the existing achievable secrecy rates in \cite{rezki2012}, which focus on passive eavesdropping attacks only. Note that our adversary model covers the possibility of a full eavesdropping attack as well since it allows for the adversary to eavesdrop (or jam) for an arbitrary fraction of the time.
\item We bound the secrecy capacity 
when there are multiple hybrid adversaries. The challenge in bounding the secrecy capacity for multiple adversaries scenario stems from the fact that, when an adversary jams the legitimate receiver, it also interferes to the other adversaries as well.  However, we show that the impact of the interference of one adversary to another adversary does not appear in the bounds, which results in a tighter upper bound. Furthermore, the bounds we provide are valid for the cases in which the adversaries collude or do not collude.  In the non-colluding case, we show that the secrecy capacity bounds are determined by the adversary that has the strongest eavesdropper channel.
\end{itemize}
In addition to  the aforementioned set-up, we also consider a delay limited communication in which a message of fixed size arrives at the encoder at the beginning of each block, and it needs to be transmitted reliably and securely by the end of that particular block. Otherwise, \emph{secrecy outage} occurs at that block. We analyze delay limited capacity subject to a secrecy outage constraint. We employ a time sharing strategy in which we utilize a portion of each block to generate secret key bits and use these key bits as a supplement to secure the delay sensitive messages that are transmitted in the other portion of each block. Our scheme achieves  positive delay limited secrecy rates whenever the secrecy capacity without any delay constraint is positive.
\subsubsection*{Related Work}
The wiretap channel, introduced by Wyner~\cite{wyner1975}, models information theoretically secure message
transmission in a point-to-point setting, where
a passive adversary eavesdrops the communication between two legitimate nodes by wiretapping the legitimate
receiver. While attempting to decipher the message, no limit
is imposed on the computational resources available to the
eavesdropper. This assumption led to defining (weak) secrecy
capacity as the maximum achievable rate subject to zero mutual
information rate between the transmitted message and the
signal received by the adversary.
This work was later generalized to the non-degraded scenario~\cite{csiszar1978} and the Gaussian channel~\cite{leung1978}. By exploiting the stochasticity and the asymmetry of wireless channels, the recent works~\cite{liang2008,lai2008} extended the results in \cite{wyner1975,csiszar1978,leung1978} to a variety of scenarios involving fading channels.
However, all of the mentioned works consider a passive adversary that can only eavesdrop.

There is a recent research interest on hybrid adversaries that can either jam or eavesdrop \cite{amariucai2012,mukher2013,zhou2012}. In~\cite{mukher2013}, 
the authors formulate the wiretap channel as a two player zero-sum game in which the payoff function is an achievable ergodic secrecy rate. The strategy of 
the transmitter is to send the message in a full power or to utilize some of the available power to produce artificial noise. The conditions under which pure 
Nash equilibrium exists are studied. In~\cite{amariucai2012}, the authors consider fast fading main and eavesdropper channels and a static jammer channel, where
the adversary follows an ergodic strategy such that it jams or eavesdrop with a certain probability in each channel use.
Under this configuration, they propose a novel encoding scheme, called block-Markov Wyner secrecy encoding. In~\cite{zhou2012}, the authors introduce a pilot contamination attack 
in which the adversary jams during the reverse training phase to prevent the transmitter from estimating the channel state correctly.  
The authors show the impact of the pilot contamination attack on the secrecy performance. 
Note that, neither of these works consider an adversary that has an \emph{arbitrary strategy} to either jam or eavesdrop, which is the focus of this paper.


Channels under arbitrary jamming (but no eavesdropping) strategies have been studied in the context of
arbitrary varying channel (AVC). AVC, the concept of which is introduced in~\cite{blackwell1960},  is defined to be the communication channel the statistics of which change in an arbitrary and unknown manner during the transmission of information.
In~\cite{csiszar1991}, the authors derive the capacity for Gaussian AVCs, memoryless Gaussian channels disrupted by a jamming signal that changes arbitrarily with unknown statistics. An extensive treatment of AVCs, outlining the challenges and existing approaches can be found in~\cite{lapidoth1998}.
Recently, discrete memoryless AVCs with a secrecy constraint and no receiver feedback have been studied in \cite{molavianjazi2009} \cite{bjelakovic2012}, where the states of the channels to the both receiver and the eavesdropper remain unknown to the legitimate pair and change in an arbitrary manner under the control of the adversary. The achievable secrecy rates they propose are zero when the worst possible transmitter-to-receiver channel is a degraded version of the best possible transmitter-to-adversary channel. On the other hand, in addition to the jamming signal of the adversary, we consider the fading channels whose states cannot be completely controlled by the adversary. We show the \emph{secrecy capacity} is zero when the main channel gain is stochastically dominated by the eavesdropper channel gain. Furthermore, under arbitrarily small receiver feedback rate (1-bit at the end of each block), we show that the secrecy capacity is \emph{non-zero}.

The rest of this paper is organized as follows. In Section~\ref{chap:system}, we explain the system model.
In Section~\ref{chap:result}, we present the secrecy capacity bounds for the no feedback case, and 
in Section~\ref{chap:1bit}, we consider the $1$-bit feedback case. In Section~\ref{sec:multi_adv}, we study the multiple adversaries case. In Section~\ref{delay_limited_sec}, we present our results related to the strict delay setting. 
In Section~\ref{chap:numeric}, we present our numerical results and conclude the paper in Section~\ref{chap:conc}.
\section{System Model }
\label{chap:system}

We study the communication system illustrated in Figure~\ref{fig:sysmdl}. In our system a transmitter has a message $w\in\mathcal{W} $ to transmit to the receiver over the main channel. The adversary chooses to either jam the receiver over the jammer channel or eavesdrop it over the eavesdropping channel. The actions of the adversary is parametrized by the state, $\phi(i)$ of a switch, shown in Figure~\ref{fig:sysmdl}. Thus, our system consists of three channels: main, eavesdropper and jammer channels, all of which are block fading.
In the block fading channel model, time is divided into discrete blocks each of which contains $N$ channel uses. 
The channel states are assumed to be constant within a block and vary independently from one block to the next. 
We assume the adversary is half duplex, i.e., the adversary can not jam and eavesdrop simultaneously.
The observed signals at the legitimate and the adversary in $i$-th block are as follows:
\begin{align}
Y^N(i)& = G_m(i)x^N(i)+ G_z(i)S_j^N(i) \phi(i)+ S_m^N(i) \label{mainCh}\\
Z^N(i)& =
\begin{cases} 
G_e(i)x^N(i) + S_e^N(i) &\mbox{ if } \phi(i)=0 \label{eavesCh}\\
\emptyset & \mbox{ if } \phi(i) =1
\end{cases}
\end{align}
\noindent where $x^N(i)$  is the transmitted signal, 
$Y^N(i)$  is the signal received by the legitimate receiver,  $Z^N(i)$ is the signal received by the adversary, $S_j^N(i)$, $S_m^N(i)$, and $S_e^N(i)$ are noise vectors distributed as complex Gaussian, $\mathcal{CN}(\mathbf{0},P_jI_{N\times N})$, $\mathcal{CN}(\mathbf{0},I_{N\times N})$, and $\mathcal{CN}(\mathbf{0},
I_{N\times N})$, respectively, and $P_j$ is the jamming power. Indicator function $\phi(i)=1$ if the adversary is in a jamming state in $i$-th block; otherwise,  $\phi(i)=0$. Channel gains, $G_m(i)$, $G_e(i)$, and $G_z(i)$ are defined to be the complex gains of the main channel, eavesdropper channel, and jammer channel, respectively (as illustrated in Figure~\ref{fig:sysmdl}). Associated power gains are  denoted with $H_m(i)= \lvert G_m(i)\rvert^2$, $H_e(i)= \lvert G_e(i)\rvert^2$, and $H_z(i)= \lvert G_z(i)\rvert^2$. For any  integer $M>0$, the joint probability density function (pdf) of $\left(G^M_m, G^M_e, G^M_z\right)$ is
\begin{align}
&p_{G^M_m, G^M_e, G^M_z} \left(g^M_m, g^M_e, g^M_z\right)\\
&\qquad\qquad\qquad\quad = \prod_{i=1}^M p_{G_m,G_e,G_z}\left(g_m(i), g_e(i),g_z(i)\right).
\end{align}
 Here, $g_m(i)$, $g_e(i)$, and $g_z(i)$ are the realizations of $G_m(i)$, $G_e(i)$, and $G_z(i)$, respectively. 
We assume that the joint pdf of instantaneous channel gains, $p_{G_m,G_e,G_z}(g_m,g_e,g_z)$ is known by all entities.
The transmitter does not know the states of any channel, and also cannot observe the strategy of the adversary in any given block. The adversary and the receiver know  $g_e(i)$ and  $(g_m(i)$, $g_z(i))$, respectively at the end of block $i$. The receiver can observe the instantaneous strategy of the adversary, $\phi(i)$ in block $i$  (e.g., via obtaining the presence of jamming) only at the end of block $i$. We generalize some of our results to the case in which the receiver cannot observe $g_z(i)$.

We consider two cases in which feedback from the receiver to the transmitter is not available or some limited feedback is available. In particular, in the latter case, we consider a 1-bit feedback over an error-free public channel 
at \emph{the end of each block}. We denote the feedback sent at $j$-th time instant as $k(j)$.

For the 1-bit feedback case, $k(j)$ is an element of  $\{0,1\}$ and is a function of $(y^{j},g_m^{i},g_z^{i}, \phi^{i})$ if time instant $j$ corresponds to the end of a block, i.e., $j=iN$ for any block index $i\geq 1$. For other time instants, the receiver does not send feedback: $k(j)=\emptyset$ if $j\neq iN$ for all $i\geq1$. For the no feedback case, $k(j)=\emptyset$ for all $j\geq 1$. 

The transmitter encodes  message $w$ over $M$ blocks. The transmitted signal at $j$-th instant, $x(j)$ can be written as 
\begin{equation}
x(j) = f_j(w, k^{j-1}),
\end{equation}
where $f_j$ is the encoding function used at time $j$.
We assume the input signals satisfy an average power constraint such that 
\begin{equation}
\frac{1}{NM}\sum_{j=1}^{NM} \mathbb E\left[\left|f_j\left(w, K^{j-1}\right)\right|^2\right] \leq P_t \label{pw_cst}
\end{equation}
for all $w\in \mathcal{W}$, where $\mathcal{W}$ is the message set. Here, the expectation is taken over $K^{j-1}=\left[K(1),\dots,K(j-1)\right]$, where $K(j)$ is the random variable denoting the feedback signal sent at $j$-th instant . 
The channels depicted in Figure~\ref{fig:sysmdl} are memoryless i.e.,

\begin{align}
&p\left(y^{N}(i),z^{N}(i)|x^{Ni},g_m^i, g_e^i, g_z^i, k^{N(i-1)},\phi^i\right)\nonumber\\
&\quad= p\left(y^N(i),z^N(i)|x^N(i), g_m(i), g_e(i), g_z(i),\phi(i)\right) \label{c1pdf} \\
&\quad= p\left(y^N(i)|x^N(i), g_m(i), g_z(i),\phi(i)\right)\times\nonumber \\
&\qquad\qquad\qquad\qquad\qquad p\left(z^N(i)|x^N(i), g_e(i),\phi(i)\right),\label{cpdf}
\end{align}
where \eqref{c1pdf} follows form the memoryless property and \eqref{cpdf} follows from the fact that the additive noise components in $y^N(i)$ and $z^N(i)$ are independent. 
Adversary strategy $\phi(i)$ changes arbitrarily from one block to the next. Here, the conditional pdfs $p\left(y^N(i)|x^N(i), g_m(i), g_z(i),\phi(i)\right)$ and
$ p\left(z^N(i)|x^N(i), g_e(i),\phi(i)\right)$ are governed by the signal models of the main channel \eqref{mainCh} and the eavesdropper channel \eqref{eavesCh}, respectively.

The transmitter aims to send message $w\in\mathcal{W}=\{1,2,\ldots2^{NMR_s}\}$ to the receiver over
 $M$ blocks with rate $R_s$. By employing a $\left(2^{NMR_s}, NM\right)$ code, the encoder at the transmitter maps message
 $w$ to a codeword $x^{NM}$, and the decoder at the receiver, $d(\cdot)$ maps the received sequence $Y^{NM}$ to $\hat{w}\in \mathcal{W}$. 
The average error probability of a $\left(2^{NMR_s}, NM\right)$ code is defined as 
\begin{equation}
P_e^{NM} = 2^{-NMR_s}\sum_{w\in\mathcal{W}} \mathbb P\left(d\left(Y^{NM}\right)\neq w| w \text{ was sent}\right)
\end{equation}

The secrecy of a transmitted message, $w$ is measured by the equivocation rate at the adversary, which
is equal to the entropy rate of the transmitted message conditioned on the adversary's observations.

\begin{definition}
A secrecy rate $R_s$ is said to be achievable if, for any $\epsilon > 0$, there exists a sequence of length 
$NM$ channel codes and sets $\mathcal{A}_M$ for which the following are satisfied 
under any strategy of the adversary, $\phi^M$:
\begin{align}
&P_e^{NM} \leq \epsilon, \label{cond1}\\
&\frac{1}{MN} H\left(W| Z^{MN}, K^{MN},g^M,\phi^M \right) \geq R_s-\epsilon,  \label{cond2}
\end{align}
for sufficiently large $N$ and $M$ and for any $g^M =\left[g_m^M ,g_{e}^M, g_z^M\right] \in \mathcal{A}_M$ such that $\mathbb P[\mathcal{A}_M]\geq 1-\epsilon$. 
\end{definition}
Note that $K^{MN}= \emptyset$ for the no feedback case. The secrecy capacity is defined to be the supremum of the achievable rates. The secrecy capacities for the no feedback and 1-bit feedback case are denoted as $C_s$ and $C^{\text{1-bit}}_s$, respectively.
Our goal is to find secrecy rates, $R_s$ that are achievable under any strategy of the adversary and find the cases in which they are tight.
\section{No Feedback}
\label{chap:result}
In this section, we provide bounds to the secrecy capacity for the no feedback case and evaluate the capacity for special cases. In the sequel, we provide a number of remarks under which we provide the basic insights drawn from the results. 
\begin{theorem}\label{thm:nocsi}\textbf{(Secrecy capacity bounds for the no feedback case)}
 The secrecy capacity, $C_s$ is bounded by
\begin{equation}
C_s^{-}\leq C_s \leq C_s^{+}
\end{equation}
where 
\begin{align}
\hspace{-1cm}C_s^{-}&= \left[\mathbb E\left[\log\left(1+\frac{P_tH_m}{1+P_jH_z}\right)-\log\left(1+P_tH_e\right)\right]\;\right]^{+}\label{nofeedbackyx}\\
C_s^{+}&= \min_{p_{\tilde{H}_m,\tilde{H}_e,\tilde{H}_z}}\mathbb E\left[\left(\log\left(1+\frac{P_t\tilde{H}_m}{1+P_j\tilde{H}_z}\right)  -\log\left(1+P_t\tilde{H}_e\right)\right)^{+}\right]\label{upp}\\
 & \text{subject to} : p_{\tilde{H}_m,\tilde{H}_z} = p_{H_m,H_z},~p_{\tilde{H}_e}=p_{H_e}\nonumber
\end{align}\qed
\end{theorem}
Notice that in Theorem~\ref{thm:nocsi}, the positive operator,\footnote{$\left[x\right]^{+}=\min(0,x).$} $\left[\cdot\right]^{+}$ is outside  the expectation in the lower bound, whereas it is inside the expectation in the upper bound. In the upper bound, minimization is over the all possible joint pdfs, $p_{\tilde{H}_m,\tilde{H}_e,\tilde{H}_z}$ that satisfy the following constraints 
$ p_{\tilde{H}_m,\tilde{H}_z} =p_{H_m,H_z}$ and $p_{\tilde{H}_e}=p_{H_e}$. Here, there is no constraint on the dependency of $\left(\tilde{H}_m,\tilde{H}_z\right)$ and $\tilde{H}_e$. 
Note that if $P_j=0$ in Theorem~\ref{thm:nocsi}, then new bounds  are valid for the scenario in which the adversary always eavesdrops the main channel, which is a common scenario in the literature. 
\begin{figure}[t]
   \centering
   \includegraphics[width=0.5\textwidth]{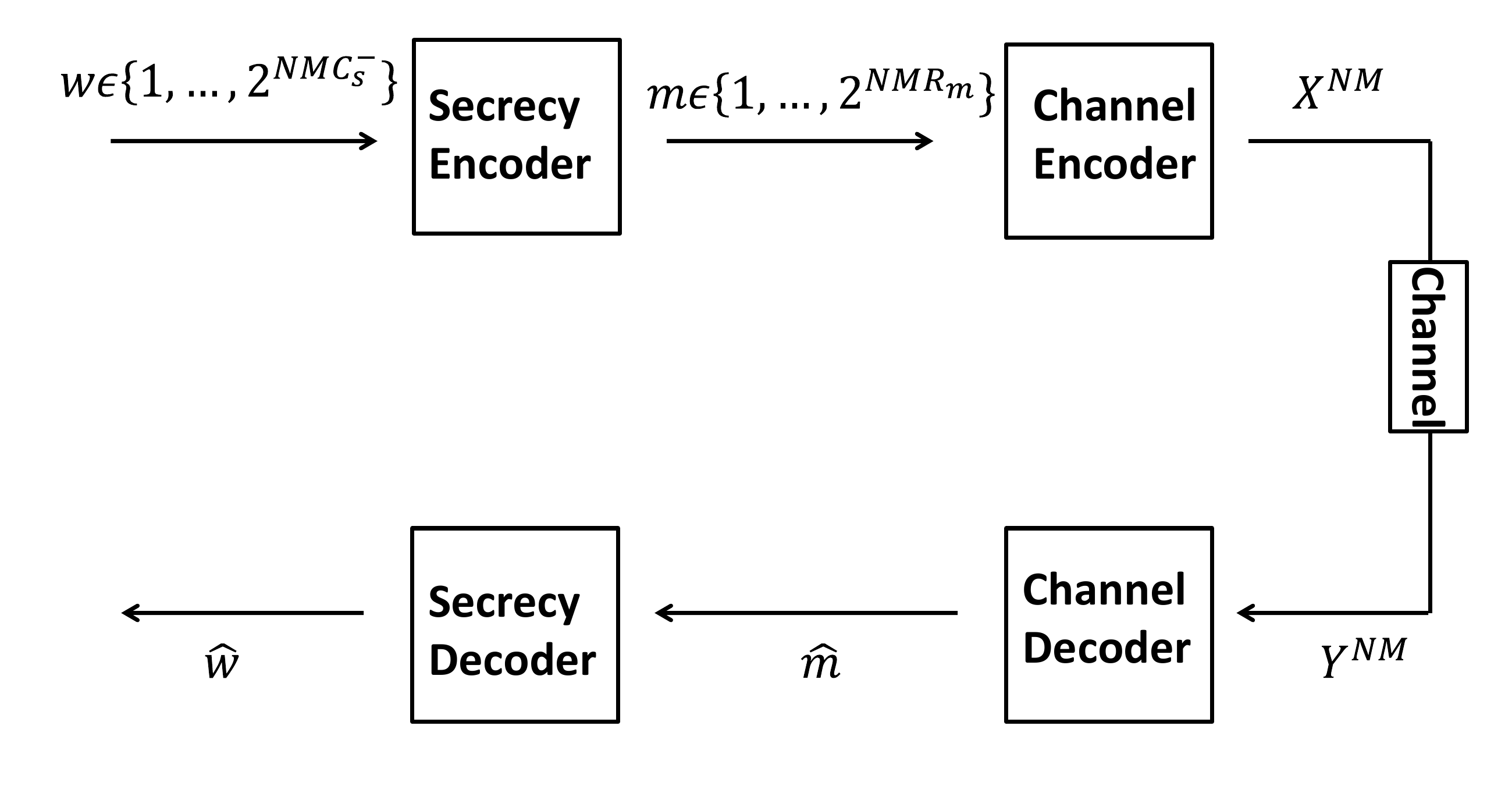}
   \caption{Achievability strategy described in the proof sketch of Theorem~\ref{thm:nocsi}.}
   \label{fig:mainmodel}
 \end{figure}

The complete proofs for the lower bound and the upper bound in Theorem~\ref{thm:nocsi} are available in Appendix~\ref{section:nocsi}. Here, we provide a proof sketch for the lower bound.
We consider the impact of the
adversary's arbitrary strategy on both the probability error and
secrecy. The secrecy encoder, depicted in Figure~\ref{fig:mainmodel}, maps message $w\in \{1,\dots,2^{NMC^{-}_s}\}$ to randomized message $m\in \{1,\dots,2^{NMR_m}\}$ as in \cite{wyner1975}. The channel encoder, illustrated in Figure~\ref{fig:mainmodel}, employs codebook $\mathit{c}$, where the codebook contains $2^{NMR_m}$ independently and identically generated codewords, $x^{NM}$ of length $NM$.  The channel encoder maps randomized message $m$ to one of the codewords in $\mathit{c}$. The probability law of the main channel is $p\left(y^N(i)|x^N(i), g_m(i), g_z(i),\phi(i)\right)$, where $\phi(i)$ changes from one block to the next arbitrarily. To remove the arbitrary nature of the main channel, the decoder artificially generates a noise sequence drawn from $\mathcal{C}\mathcal{N}\left(0, h_zP_jI_{N\times N}\right)$, where $h_z$ is picked from $H_z(i)$, and adds the noise sequence to it's received signal $y^N(i)$
when the adversary is in the eavesdropping state, $\phi(i)=0$. Hence, the
decoder can employ typical set decoding~\cite{cover1991}, which would not have been possible without the artificial noise, due to the lack of the underlying probability distribution for the received signal associated with the arbitrary adversary strategy. 
We select $R_m$ as
\begin{align}
R_m&=\max_{p(x^N(i))} \frac{1}{N}I(X^N(i), Y^N(i)| G_m(i), G_z(i),\phi(i) =1)\label{xcxc}\\
&=\mathbb E\left[\log\left(1+\frac{P_tH_m}{1+P_jH_z}\right)\right], \label{xcxc1}
\end{align} 
where the joint distribution of $\left(X^N(i), Y^N(i)\right)$ in \eqref{xcxc} is governed by \eqref{mainCh} for a given $p(x^N(i))$, and~\eqref{xcxc1} follows from the fact that $X^N(i) \sim \mathcal{C}N (0, P_tI_{N\times N})$ maximizes the optimization in \eqref{xcxc}. Therefore, each codeword, $x^{NM}$ is picked from  $\mathcal{C}N (0, P_tI_{NM\times NM})$.  For the equivocation analysis, the possibility of the adversary eavesdropping at all times should be taken into account. We need to use a conservative secrecy encoder, designed for $\phi(i)=0$ for all $i\geq1$; otherwise, we cannot achieve an arbitrarily low mutual information leakage rate to the adversary with high probability. With the aforementioned techniques, we show that $C_s^{-}$ satisfies constraints \eqref{cond1} and \eqref{cond2} in Appendix~\ref{section:nocsi}. We now provide several remarks related to Theorem~\ref{thm:nocsi}.

\begin{remark}\textbf{(Secrecy capacity is zero when the eavesdropper channel power gain stochastically dominates the main channel effective power gain)}
\label{dom_remark}
If $H_e$ stochastically dominates\footnote{Random variable $A$ stochastically dominates random variable $B$ if 
$
F_A(a) \leq F_B(a) \text{  for all  } a $,
where $F_A(a) \triangleq \mathbb P[A\leq a]$ and $F_B(a) \triangleq \mathbb P[B\leq a]$ .} the main channel effective power gain, $H_m^{*}\triangleq\frac{H_m}{1+P_jH_z}$, and $H_e$ and $H^{*}_m$ have continuous cumulative distribution functions (cdfs), then the secrecy capacity, $C_s$ is zero. To observe this fact, let $\hat{H}_e\triangleq F_{H_e}^{-1}\left(F_{H_m^{*}}\left(H_m^{*}\right)\right)$, where $F_A$ and $F_A^{-1}$ stand for the cdf and the inverse cdf\footnote{Inverse cdf is generally defined as $F_A^{-1} (a) \triangleq\inf \left\{b: F_A(b)\geq a\right\}$. However, since we assume that $A$ has a continuous cdf, $F_A^{-1} (a) \triangleq\inf \left\{b: F_A(b)= a\right\}$. } of random variable $A$, respectively. From the definition of stochastic dominance and the definition of $\hat{H}_e$, we have $\hat{H}_e \geq H_m^{*} $ with probability 1. We now show that $\hat{H}_e$ and $H_e$ have the same cdf  the following derivation: 
\begin{align}
\mathbb P\left[\hat{H}_e \leq a\right] &= \mathbb P\left[F_{H_e}^{-1}\left(F_{H_m^{*}}\left(H_m^{*} \right)\right) \leq a\right]\\
&=\mathbb P\left[F_{H_m^{*} }(H_m^{*} )\leq F_{H_e}(a)\right]\label{rem1}\\
&=\mathbb  P\left[H_m^{*} \leq F^{-1}_{H_m^{*} }\left(F_{H_e}\left(a\right)\right)\right] \label{rem2}\\
&=F_{H_m^{*} }\left(F^{-1}_{H^{*}_m}\left(F_{H_e}\left(a\right)\right)\right) \\
&=F_{H_e}(a), \forall a\geq 0, \label{rem3}
\end{align}
where \eqref{rem1} follows from the fact that $F_A(c) \leq b \iff c \leq F^{-1}_A(b)$ with $b\in \left[0,1\right]$, and \eqref{rem2} and \eqref{rem3} follow from the continuity of the cdf of $H_m^{*}$.  Hence, $(H_m, \hat{H}_e, H_z)$ satisfy the constraint given in the upper bound~\eqref{upp}. When $(\tilde{H}_m, \tilde{H}_e, \tilde{H}_z)=(H_m, \hat{H}_e, H_z)$, the expectation term in \eqref{upp} is zero.\qed
\end{remark}

Remark~\ref{dom_remark} is easy to state for the fading scenario in which $H_m$ and $H_e$ are exponentially distributed random variables. Condition $\mathbb E[H_m] \leq \mathbb E[H_e]$ is sufficient for $H_e$ to stochastically dominate  $H_m^{*}$ (defined in Remark~\ref{dom_remark}).

\begin{remark}\textbf{(Bounds are tight if the power gain of the effective main channel  is larger than that of the eavesdropper channel with probability 1)}
Suppose there exits random variables $\hat{H}_m,\hat{H}_e$, and $\hat{H}_z$ satisfying the following conditions:1)
\begin{equation}
\frac{\hat{H}_m}{1+P_j\hat{H}_z} \geq  \hat{H}_e   \label{capacity_cond}
\end{equation}
with probability 1,  2) $p_{\hat{H}_m,\hat{H}_z} = p_{H_m,H_z}$, and 3) $p_{\hat{H}_e} = p_{H_e}$.  Then, $C_s^{-}= C_s = C_s^{+}$. To observe this fact, let  $(\tilde{H}_m, \tilde{H}_e, \tilde{H}_z)$ in \eqref{upp} be $(\hat{H}_m, \hat{H}_e, \hat{H}_z)$. Then, the positive operator gets out of the expectation in the upper bound, $C_s^{+}$.  Furthermore, since the lower bound does not depend on $p_{H_m,H_e,H_z}$ but depend on $p_{H_m,H_z}$ and $ p_{H_e}$, we can replace $(H_m, H_e, H_z)$ with  $(\hat{H}_m, \hat{H}_e, \hat{H}_z)$. Thus, the upper and lower bounds become equal. \qed
\end{remark}
\begin{remark}(\textbf{When the jamming channel gain is not available at RX, the lower bound decreases.})
In Theorem~\ref{thm:nocsi}, the receiver is assumed to know $g_z(i)$. Now, suppose that the receiver is kept ignorant of $g_z(i)$. Then, the following rate 
\begin{equation}
R^{'}_s= \left[R -\mathbb E\left[\log\left(1+P_tH_e\right)\right]\;\right]^{+}
\end{equation}
is achievable,  where 
\begin{equation}
R =
\max_{p_{X^N(i)}(x^N(i))} \frac{1}{N}I(X^N(i), Y^N(i)| G_m(i),\phi(i) =1).
\end{equation}
Here, for a given $p_{X^N(i)}(x^N(i))$, the joint distribution of $\left(X^N(i), Y^N(i)\right)$ is governed by \eqref{mainCh}. We can lower bound $R$ with the following steps:
\begin{align}
R &\geq \frac{1}{N}I(X_G^N(i), Y^N(i)| G_m(i),\phi(i) =1) \label{flb}\\
    &\geq \mathbb E\left[\log\left(1+\frac{P_tH_m}{1+P_j\mathbb E[H_z]}\right)\right],\label{slb}
\end{align}
where  $X_G^N(i) \sim \mathcal{C}\mathcal{N}(0,P_tI_{N\times N})$ in \eqref{flb}. The covariance matrix of the jamming component in $Y^N(i)$ is $\mathbb E\left[H_z(i)\right]I_{N\times N}$. In~\cite{diggavi2001}, the authors show that  Gaussian noise that has the same covariance matrix with the original additive noise component minimizes $I(X^N(i); Y^N(i))$ when $X^N(i)$ is Gaussian distributed. Hence, we replace  $G_z(i)S_j^N(i)$ with $\mathcal{C}\mathcal{N}(0,\mathbb E\left[H_z(i)\right]I_{N\times N})$, and reach the inequality in \eqref{slb}.\qed
\end{remark}
Suppose that the transmitter and the adversary power constraints scale in the same order, parametrized by $P$, i.e.,  $P_t\left(P\right)= \mathcal{O}\left(P_j(P)\right)$ as $P\to\infty$. We show that the secrecy capacity is zero in the no feedback case as $P\to\infty$ in the following corollary.
\begin{corollary} (\textbf{Secrecy capacity goes to zero when the jamming and transmission power constraints scale similarly.})\label{asym_power} Suppose that $P_t(P)$ and $P_j(P)$ are continuous functions of  $P$ with $\lim_{P\to \infty} P_t(P)=\infty$, $\lim_{P\to \infty} P_j(P)=\infty$ and $P_t\left(P\right)= \mathcal{O}\left(P_j(P)\right) \text{ as } P\to\infty$.
When the power gains of the channels have bounded and continuous  pdfs  and have finite expectations, the secrecy capacity of the no feedback case, $C_s$ is asymptotically
\begin{equation}
\lim_{P\to\infty} C_s= 0.\label{asmNoCSI}
\end{equation}\qed
\end{corollary}

The proof of Corollary~\ref{asym_power} is available at Appendix~\ref{high_power_proof}. To prove~\eqref{asmNoCSI}, we investigate  the upper bound, $C_s^{+}$ as $P\to \infty$ and show that 
\begin{equation}
\lim_{P \to \infty} C^{+}_s = 0. \label{asm2NoCSI}
\end{equation}

\section{1-Bit Feedback}
\label{chap:1bit}
In this section, we analyze the secrecy capacity for the 1-bit feedback case, i.e., the receiver is allowed to send  a 1 bit feedback over a public channel at the end of each block. As we observe in Remark~\ref{dom_remark},  the secrecy capacity of the no feedback case is zero if $H_e$ stochastically dominates $H^{*}_m\triangleq\frac{H_m}{1+P_jH_z}$. However, in this section, we show that the lower bound for the 1-bit feedback case is non-zero for the most of the joint pdfs of power gains.
\begin{theorem}(\textbf{Secrecy capacity bounds for the 1-bit feedback case})\label{1bitrate_accum}
The secrecy capacity, $C_s^{\text{1-bit}}$  is bounded by 
\begin{equation}
\max\left(C_s^{-},R_s^{\text{1-bit}} \right)\leq C_s^{\text{1-bit}} \leq C_s^{+\text{1-bit}}
\end{equation}
where
\begin{align}
& C_s^{+\text{1-bit}}= \mathbb E\left[\log\left(1+\frac{P_tH_m}{1+\max(P_jH_z, P_tH_e)}\right)\right]\\
&R_s^{\text{1-bit}} =  \max_R\frac{1}{\mathbb E[T]} \mathbb E\left[R- \log\left(1+P_t\sum_{i=1}^{T} \tilde{H}_{e}(i)\right) \right]^{+} \label{1bit_accum}
\end{align}
where $C_s^-$ is provided in~\eqref{nofeedbackyx},
$T$ is a random variable with probability mass function (pmf), $p_{T}(t)=\mathbb P(D_t \cap D_{t-1}^c)= \mathbb P(D_t)-\mathbb P(D_{t-1})$, $t\geq 1$ with $D_t \triangleq \left \{ \log\left(1+\sum_{i=1}^t\frac{P_tH_m(i)}{1+P_jH_z(i)}\right) \geq R \right\}$ and $D_0 = \emptyset$, and
\begin{align}
&p_{\tilde{H}_e(1), \tilde{H}_e(2),\dots, \tilde{H}_e(T)|T}\left(h_e(1),h_e(2),\dots,h_e(T)|T=t\right)=\nonumber\\
&p_{H_e(1),H_e(2),\ldots, H_e(t)}\left(h_e(1),h_e(2),\dots,h_e(t)|D_t, D^c_{t-1}\right)\nonumber
\end{align}\qed
\end{theorem}
The complete proofs for lower and upper bounds are available in Appendix ~\ref{app:ach_csi}. Note that the feedback available at the transmitter in block $i$, $K^{(i-1)N}$ is independent from the channel gains in block $i$, $G(i)$ since the transmitter observes the feedback at the end of the block, and the channel gains change from one block to the next independently. Hence,  the transmission power term in the upper bound is not a function of the channel gains and is equal to the transmission power constraint, 
$P_t$. Furthermore, notice that in Theorem~\ref{1bitrate}, the positive operator is inside the expectation in \eqref{1bit}, that makes the lower bound positive for a wide class of channel statistics. 

\begin{remark}(\textbf{Non-zero secrecy capacity})\label{nonzero}  Note that $\left\{\log\left(1+\frac{P_tH_m(i)}{1+P_jH_z(i)}\right)\right\}_{i\geq 1}$ is a sequence of i.i.d non-negative random variables and $T= \inf\left\{t:\sum_{i=1}^{t} \log\left(1+\frac{P_tH_m(i)}{1+P_jH_z(i)}\right)\geq R\right\}$. If $\mathbb P\left[\frac{P_tH_m}{1+P_jH_z}\neq 0\right]>0$, $\mathbb E[T] <\infty$ for all $R>0$~\cite{ross1995}. Furthermore, there exists $R\geq 0$ that makes  $\mathbb E\left[R- \log\left(1+P_t\sum_{i=1}^{T} \tilde{H}_{e}(i)\right) \right]^{+}$ also positive since $\mathbb P\left[ \sum_{i=1}^{T} \tilde{H}_{e}(i)<\infty\right] >0$. Hence, we observe that $C_s^{\text{1-bit}} >0$.\qed
\end{remark}

\begin{figure}[t]
   \centering
   \includegraphics[width=0.5\textwidth]{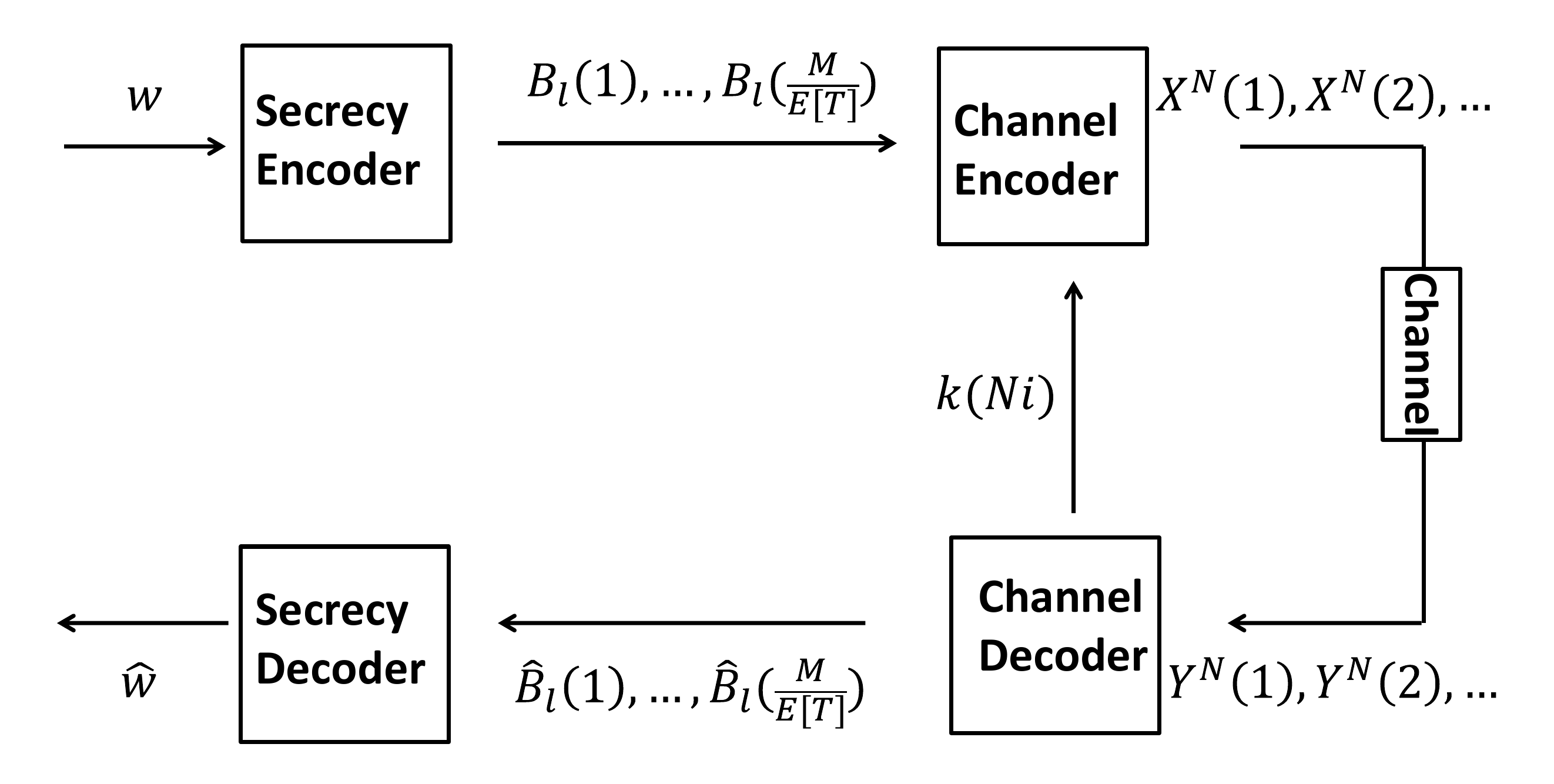}
   \caption{Achievability strategy described in the proof sketch of Theorem~\ref{1bitrate_accum}. The feedback at at the end of block $i$ is denoted as $k(Ni)$,  where $N$ is the length of a block.}
   \label{fig:feedbackmodel}
 \end{figure}

Here, we provide the proof sketch of the lower bound provided in Theorem~\ref{1bitrate}. First, $C_s^{-}$ is achieved with the strategy provided in Theorem~\ref{thm:nocsi} without the feedback. The strategy to achieve $R_s^{{\text{1-bit}}}$ is as follows. The secrecy encoder, depicted in Figure~\ref{fig:feedbackmodel}, maps message $w\in\left[1:2^{NMR_s^{{\text{1-bit}}}}\right]$ to bit sequence $B_l\in \{0,1\}^{NM\frac{R}{\mathbb E[T]}}$ of size $NM\frac{R}{\mathbb E[T]}$ with a stochastic mapping as described in ~\cite{wyner1975}, where $l\in [1,2,\dots,2^{NM\frac{R}{\mathbb E[T]}}]$. Bit sequence $B_l$ is partitioned into the bit groups $\{B_l(k)\}_{ k\in \left[1,2,\dots,\lceil \frac{M}{\mathbb E[T]}\rceil\right]}$ each of which has size of $NR$ bits such that $B_l = [B_l(1), B_l(2),\ldots, B_l(\lceil \frac{M}{\mathbb E[T]}\rceil)]$. The channel encoder, depicted in Figure~\ref{fig:feedbackmodel}, generates  Gaussian codebook $\mathit{c}$ of size $2^{NR}$, and each bit group $B_l(k)$ is mapped to one of the codewords in the codebook.

To send $B_l(k)$ in block $i$, the associated codeword $x^N(i)$ is transmitted over the channel. The channel encoder keeps sending the same codeword until $B_l(k)$ is successfully decoded. The channel decoder, depicted in Figure~\ref{fig:feedbackmodel}, employs maximum ratio combining (MRC), and combines all received sequences associated with $B_l(k)$. Specifically, the channel decoder multiples each $y^N(i)$ associated with the bit group with $\frac{g^{*}_m(i)}{\left(1+P_jh_z(i)\right)^2}$ and sums them. From the random coding arguments, we can see that $B_l(k)$ will be decoded with arbitrarily low probability error at $i$-th block if the event $S(i)\triangleq\left\{\log\left(1+\sum_{j=1}^{r(i)}\frac{P_tH_m(i-j+1)}{1+P_jH_z(i-j+1)}\right)\geq R\right\}$ occurs, regardless of the adversary strategy $\phi^i$, where $r(i)$ is the number of transmissions for $B_l(k)$ until the end of block $i$. If event $S(i)$ occurs, the channel decoder sends back positive acknowledgment signal (ACK), and the channel encoder sends the next bit group, i.e., $B_l(k+1)$ on block $i+1$. If event $S(i)$ does not occur, the channel decoder feeds back a negative acknowledgment signal (NAK) at the end of block $i$. On next block $i+1$, the channel encoder sends the same codeword, i.e., $x^N(i+1)= x^N(i)$. This process is repeated until $B_l$ is successfully decoded.


In the derivation for the lower bound for the equivocation rate, we assume that the adversary can observe the transmissions in the jamming state\footnote{We will drop this assumption when we analyze the case in which the transmitter has the main channel state information (CSI) in addition to the 1-bit feedback (Corollary~\ref{1bitrate_main CSI}).}. Consider a renewal process in which a renewal occurs when the accumulated mutual information associated with a bit group exceeds threshold $R$ for the first time. 
In Appendix~\ref{app:ach_csi}, we show that 1-bit feedback case can be considered as a model in which \emph{secure} bits of random size  $N\left[R- \log\left(1+P_t\sum_{i=1}^T \tilde{H}_{e}(i)\right) \right]^{+}$ are decoded successfully at each renewal point. Here, $T$ is the random variable given in Theorem~\ref{1bitrate_accum}, and it represents the number of transmissions for a bit group. Thus, $N\log\left(1+P_t\sum_{i=1}^T \tilde{H}_{e}(i)\right)$ can be considered as a random amount of accumulated mutual information at the adversary corresponding to the transmissions of a bit group. Theorem~\ref{1bitrate} follows  when we apply the renewal reward theorem~\cite{caire2001}, where the rewards are the successfully decoded secure bits at each renewal instants, i.e.,
\begin{equation}
\lim_{M\to\infty} \frac{R_{w}(M)}{MN}= \frac{1}{\mathbb E[T]} \mathbb E\left[R- \log\left(1+P_t\sum_{i=1}^T \tilde{H}_{e}(i)\right) \right]^{+}
\end{equation}
with probability 1, where $R_{w}(M)$  is defined to be the amount of secure bits (explained above) accumulated at the receiver up to block $M$.

Instead of employing MRC strategy, the receiver can employ a plain automatic repeat request (ARQ) strategy in which the receiver discards the received sequence $y^N(i)$ when the decoding error occurs on $i$-th block.  Impact of plain ARQ on the lower bound is captured with the following corollary.
\begin{corollary}(\textbf{Secrecy capacity lower bound with plain ARQ})\label{1bitrate}
The secrecy capacity, $C_s^{\text{1-bit}}$  is bounded by 
\begin{equation}
\max\left(C_s^-, R_s^{*{\text{1-bit}}} \right)\leq C_s^{\text{1-bit}}
\end{equation}
where
\begin{equation}
R_s^{*{\text{1-bit}}} =\max_R  \;p\times \mathbb E\left[R- \log\left(1+P_t\sum_{i=1}^{T^{*}} \tilde{H}_{e}(i)\right) \right]^{+}\label{1bit}
\end{equation}
where $C_s^-$ is provided in \eqref{nofeedbackyx}. In~\eqref{1bit}, $p \triangleq \mathbb P\left(\log\left(1+\frac{P_tH_M}{1+P_jH_z}\right) \geq R\right)$, $T^{*}$ is a random variable with probability mass function (pmf), $p_{T^{*}}(t)= p(1-p)^{t-1}$, $t\geq 1$, and
\begin{align}
&p_{\tilde{H}_e(1), \tilde{H}_e(2),\dots, \tilde{H}_e(T^{*})|T^{*}}\left(h_e(1),h_e(2),\dots,h_e(T^{*})|T^{*}=t\right)=\nonumber\\
&\qquad\prod_{i=1}^{t-1}p_{H_e}\left(h_e(i)|R> \log\left(1+\frac{P_tH_m}{1+P_jH_z}\right)\right)\times\nonumber\\
&\quad\qquad\qquad p_{H_e}\left(h_e(t)|R\leq \log\left(1+\frac{P_tH_m}{1+P_jH_z}\right)\right).
\end{align}\qed
\end{corollary}
The proof of Corollary~\ref{1bitrate} can be found at the end of achievability proof of Theorem~\ref{1bitrate_accum}. It can be observed that the lower bound in Corollary~\ref{1bitrate} is not larger than the lower bound in Theorem~\ref{1bitrate_accum}. 

In~\cite{rezki2012}, the authors consider a scenario in which the adversary is a fully eavesdropper, and the transmitter has no information of the states of main and eavesdropper channels, which change from one block to the next randomly as described in our scenario. For the case in which 1-bit feedback is available at the end of each block, the authors employ the plain ARQ strategy mentioned above to achieve the secrecy rate in Theorem 2 of~\cite{rezki2012}. However, in the secrecy analysis, the authors consider the impact of the bit groups, $B_l(k)$ that are successfully decoded only in a single transmission on the equivocating rate. In this paper, regardless of the number of the required transmissions for the bit groups,  we consider the impact of the each bit group on the equivocation rate with the strategy mentioned in the proof sketch of Theorem~\ref{1bitrate_accum}. Thus, we improve the achievable secrecy rate in~\cite{rezki2012} by employing  a renewal based analysis and MTC.

In Theorem~\ref{1bitrate_accum} and Corollary~\ref{1bitrate}, we observe that the information corresponding to the retransmissions of a bit group is accumulated at the adversary, which reduces the lower bound. As we will show, we can avoid this situation if the main CSI is available at the beginning of each block at the transmitter in addition to the 1-bit feedback at the end of each block. By using the rate adaptation strategy that we will introduce, the legitimate pair can ensure that information corresponding to the retransmissions of a bit group is not accumulated at the adversary.
\begin{corollary}(\textbf{Achievable secrecy rate with main CSI})\label{1bitrate_main CSI} If main CSI is available at the transmitter and the adversary, the secrecy capacity with 1-bit feedback at the end of each block is lower bounded by
\begin{equation}
 R_s^{\text{1-bit+CSI}} =\max_R  \;p\times \mathbb E\left[R- \log\left(1+P_t H_{e}\right) \right]^{+}\leq C_s^{\text{1-bit+CSI}},\label{1bit_mainCSI}
\end{equation}
where 
$p\triangleq \mathbb P\left(\log\left(1+\frac{P_tH_M}{1+P_jH_z}\right) \geq R\right)$.\qed
\end{corollary}
We omit the proof since it follows from an identical line of argument as the proof of Theorem~\ref{1bitrate_accum}. The only difference is that the legitimate pair employs a plain ARQ strategy as in Corollary~\ref{1bitrate}, and the transmitter employs a rate adaptation strategy to utilize the main CSI such that $R(i) = R$ if $R\leq \log(1+Ph_m(i))$; otherwise, $R(i)=0$, where $R$ is the rate of the Gaussian codebook used in the achievability proof of Theorem~\ref{1bitrate_accum}.  Since the transmitter keeps silent on the blocks in which condition $R> \log(1+Ph_m(i))$ is satisfied, the decoding error event occurs only when the adversary is in the jamming state. Hence, the adversary cannot hear the retransmissions because of the half duplex constraint, and  information that corresponds to the retransmissions of a bit group is not accumulated as seen in \eqref{1bit_mainCSI}.


Note that main CSI combined with 1 bit feedback provides the transmitter \emph{perfect knowledge} of the adversary jamming state (but with one block delay) since an ACK indicates that the adversary is in the eavesdropping state, and a NAK indicates that the adversary is in the jamming state in the 
previous block. Therefore, we do not need to employ a conservative secrecy encoder to account for the adversary that eavesdrops at all times.


\section{Multiple Adversaries}
\label{sec:multi_adv}
\begin{figure}[t]
   \centering
   \includegraphics[width=0.5\textwidth]{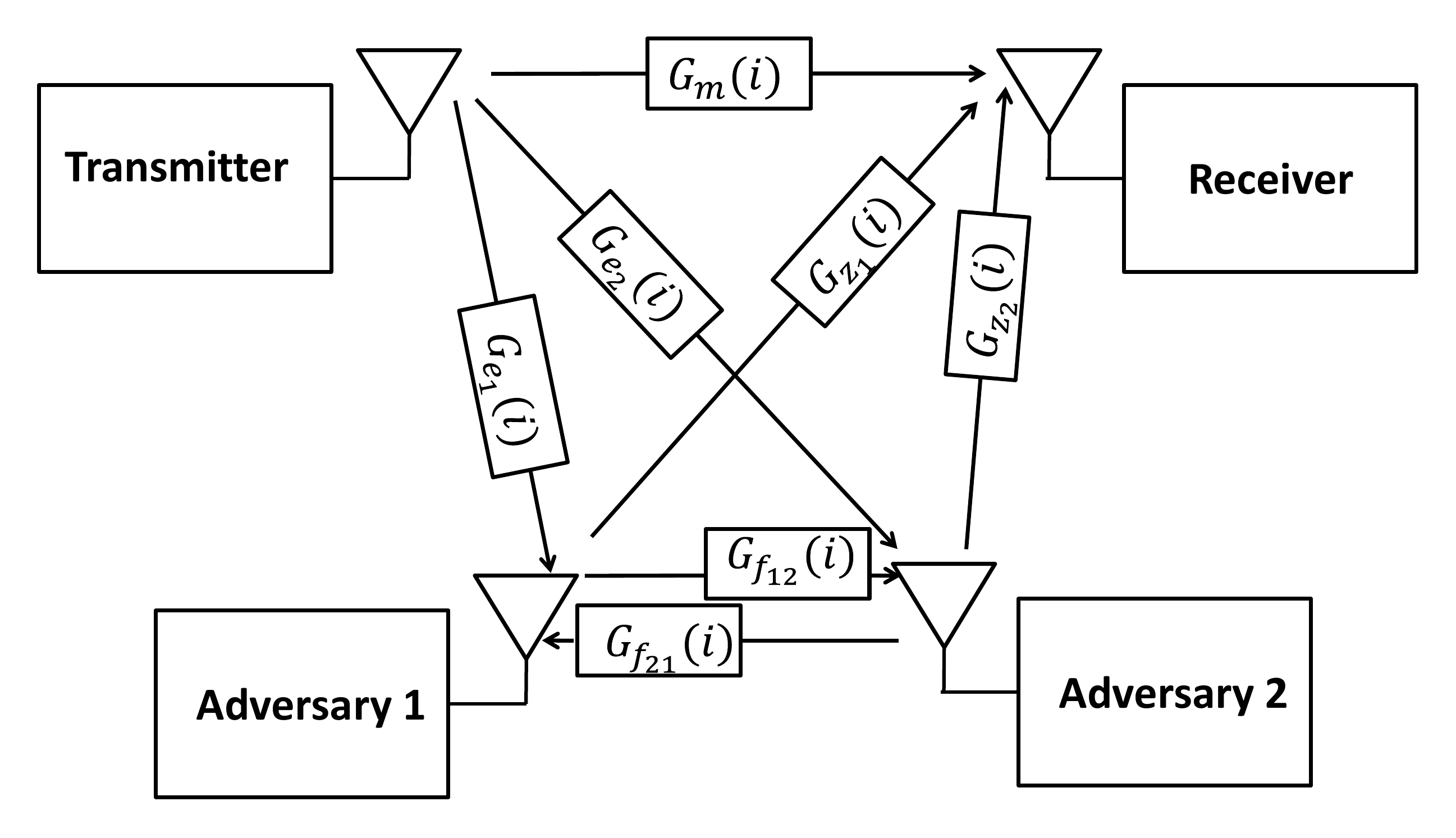}
   \caption{System model for multi-adversary scenario including two adversaries.}
   \label{fig:sysmdl2}
 \end{figure}

In this section, we  study the multiple adversary scenario in which there are $S$ half duplex adversaries each of which has an arbitrary strategy
from one block to the next. We focus on the no feedback case. The results given in this section can be extended to the 1-bit feedback case straightforwardly.   Since there are multiple adversaries, the message has to be kept secret from each adversary. Moreover, when an adversary jams the receiver, it also jams the other adversaries.  Consequently, the observed signals at the legitimate receiver and adversary $s$ in $i$-th block can be written as follows:
\begin{align}
&Y^N(i) = G_m(i)x^N(i)+ \sum_{s=1}^S G_{z_{s}}(i)S_{j_s}^N(i) \phi_s(i)+ S_m^N(i) \label{mainCh111}\\
&Z_s^N(i) =
\begin{cases}G_{e_s}(i)x^N(i)+\sum_{r=1,r\neq s}^S G_{f_{rs}}(i)S_{j_s}^N(i)\phi_r(i)  + S_e^N(i) &\mbox{ if } \phi_s(i)=0 \label{eavesCh111}\\
 \emptyset  &\mbox{ if } \phi_s(i) =1
\end{cases}
\end{align}
where $S_{j_s}$ is the jamming signal of adversary $s$, and is distributed with $\mathcal{CN}(\mathbf{0},P_jI_{N\times N})$. As depicted in Figure~\ref{fig:sysmdl2}, $G_{e_s}(i)$, $G_{z_{s}}(i)$, and $G_{f_{rs}}(i)$  are defined to be the independent complex gains of transmitter-to-adversary $s$ channel,  adversary $s$-to-receiver channel , and adversary $r$-to-adversary $s$ channel, respectively. Associated power gains are denoted with $H_{e_s}(i)= \lvert G_{e_s}(i)\rvert^2$, $H_{f_{rs}}(i)= \lvert G_{f_{rs}}(i)\rvert^2$, and $H_{z_s}(i)= \lvert G_{z_s}(i)\rvert^2$. Indicator function $\phi_s(i)=1$, if the adversary $s$ is in a jamming state in $i$-th block; otherwise,  $\phi_s(i)=0$.

For the multi adversary scenario, $\phi^M$ in \eqref{cond2} is replaced with   $\{\phi^M_s\}_{1\leq s \leq S}$, and the constraints \eqref{cond1}-\eqref{cond2} have to be satisfied for all  $\{\phi_s^M\}_{1\leq s \leq S}$. We study two types of multi-adversary scenarios: colluding and non-colluding. In the colluding scenario, the adversaries share their observations, $\left\{Z_s^{NM}\right\}$ error free whereas in the non-colluding scenario, the adversaries are not aware of the observations of each other. Hence, for the non-colluding scenario, constraint \eqref{cond2} needs to be satisfied for each adversary and for the colluding scenario, equivocation is conditioned on the adversaries' joint knowledge, i.e., $Z^{MN}$ in \eqref{cond2}
is replaced with  $\{Z_s^{MN}\}_{1\leq s \leq S}$. 
We use notations $C_s^{C}$ and $C_s^{NC}$ to denote the secrecy capacities for the colluding case and the non-colluding case, respectively. 
We first analyze the non-colluding scenario.
\begin{theorem}\textbf{(Secrecy capacity bounds for non-colluding adversaries)} \label{multi_no_csi}
The secrecy capacity of the non-colluding multiple adversary scenario, $C^{NC}_s$ under the no feedback case is bounded by
\begin{equation}
C_s^{NC-}\leq C^{NC}_s \leq C_s^{NC+}
\end{equation}
where 
\begin{align}
&C_s^{NC-}=\min_{1\leq s \leq S}\left[\mathbb E\left[\log\left(1+\frac{P_tH_m}{1+P_j\hat{H}_z}\right)-\log\left(1+P_tH_{e_s}\right)\right]\right]^{+}\\
&C_s^{NC+}= \min_{1\leq s \leq S}\quad\min_{p_{\tilde{H}_{e_1},\dots,\tilde{H}_{e_S},\tilde{H}_m,\tilde{H}_{z_1},\dots,\tilde{H}_{z_S}}} \mathbb E\left[\left(\log\left(1+\frac{P_t\tilde{H}_m}{1+P_j\tilde{H}_z}\right)-\log\left(1+P_t\tilde{H}_{e_s}\right)\right)^{+}\right]\\
&\qquad \qquad\qquad\qquad\text{subject to:  }p_{\tilde{H}_{e_1},\dots,\tilde{H}_{e_S}}=p_{H_{e_1},\dots,H_{e_S}},\;\; p_{\tilde{H}_m,\tilde{H}_{z_1},\dots,\tilde{H}_{z_S}} = p_{H_m,H_{z_1},\dots,H_{z_S}} \nonumber 
\end{align}
where $S$ is the number of the adversaries,  $\hat{H}_z \triangleq \sum_{k=1}^S H_{z_s}$,  and $\tilde{H}_z \triangleq \sum_{s=1}^S \tilde{H}_{z_s}$.\qed
\end{theorem}
The proofs of the lower and upper bounds can be found in Appendix \ref{multi_no_csi_proof}.
\begin{theorem}\textbf{(Secrecy capacity bounds for colluding adversaries)}\label{colluding}
The secrecy capacity of the colluding multiple adversary scenario, $C^{C}_s$ under the no feedback case is bounded by
\begin{equation}
C_s^{C-}\leq C^{C}_s \leq C_s^{C+}
\end{equation}
where 
\begin{align}
&C_s^{C-}=\mathbb E\left[\log\left(1+\frac{P_tH_m}{1+P_j\hat{H}_z}\right)-\log\left(1+P_t\sum_{s=1}^S H_{e_s}\right)\right]^{+}\nonumber\\
&C_s^{C+}=\min_{p_{\tilde{H}_{e_1},\dots,\tilde{H}_{e_S},\tilde{H}_m,\tilde{H}_{z_1},\dots,\tilde{H}_{z_S}}} \mathbb E\left[\left(\log\left(1+\frac{P_t\tilde{H}_m}{1+P_j\tilde{H}_z}\right)-\log\left(1+P_t\sum_{k=1}^S \tilde{H}_{e_s}\right)\right)^{+}\right]\\
&\qquad \qquad\text{subject to:  }p_{\tilde{H}_{e_1},\dots,\tilde{H}_{e_S}}=p_{H_{e_1},\dots,H_{e_S}},\;\; p_{\tilde{H}_m,\tilde{H}_{z_1},\dots,\tilde{H}_{z_S}} = p_{H_m,H_{z_1},\dots,H_{z_S}} \nonumber
\end{align}
where $S$,  $\hat{H}_z$, and $\tilde{H}_z$ are defined in Theorem~\ref{multi_no_csi}.\qed
\end{theorem}
The proof of Theorem~\ref{colluding} is similar to the proof Theorem~\ref{thm:nocsi} since the colluding scenario can be considered as a single adversary scenario, in which the adversary observes  $\{Z_s^{MN}\}_{1\leq s \leq S}$ instead of $Z_s^{NM}$. 
As seen in Theorems~\ref{multi_no_csi} and~\ref{colluding}, colluding strategy severely affects the achievable secrecy rate.
\begin{remark}\textbf{(Independence of upper bound from cross-interference)}
In~\eqref{eavesCh111}, we observe that the received signal at $s$-th adversary includes the jamming signals of the other adversaries, i.e., $\sum_{r=1,r\neq s}^S G_{f_{rs}}(i)S_{j_s}^N(i)\phi_r(i)$. We expect that these cross interference terms at the adversaries help the legitimate pair to communicate at high secrecy rates. However, as seen in Theorem~\ref{multi_no_csi} and \ref{colluding}, the upper bounds (and also lower bounds) are independent of these jamming terms. Note that the secrecy constraint in the proof of upper bounds makes the minimization of the equivocation rate over the adversary strategies  arbitrarily close to the message rate. The strategies that minimize the equivocation rate in the proofs are the ones in which all adversaries eavesdrop the main channel. Hence, the upper bound derivation becomes independent of the cross interference across the adversaries. The detailed information can be found in Appendix~\ref{multi_no_csi_proof}.\qed
\end{remark}
\section{Strict Delay}\label{delay_limited_sec}

In this section, we address  the problem with $1$-block delay constraint: At the beginning of each block $i$, $1\leq i \leq M$, message 
$w(i) \in \{1,\ldots,2^{N R_s}\}$ becomes available at the encoder, and needs to be securely communicated to the receiver by the end of block $i$.
Note that, the definition of secrecy capacity needs to be restated with the delay requirement.
We consider a set of codes of rate $R_s$ where the transmitter maps message $w(i)$, and the previously transmitted signals\footnote{Note that,
the encoded signal $x^{N}(i)$ also depends on the previously transmitted signals $\{x^{N}(j)\}_{j=1}^{i-1}$. It is required
to utilize secrecy banking argument \cite{gungor2013secrecy}, in which shared secrets are stored to be utilized in later blocks.}
 $\{x^{N}(j)\}_{j=1}^{i-1}$ to $x^{N}(i)$,
and the decoder maps the received sequence $y^{N}(i)$ to $\hat{w}(i)$. The error event is defined as
\begin{align}
E(i)\triangleq\{W(i)\neq \hat{W}(i)\}.
\end{align}
When $w(i)$ cannot be communicated reliably or securely at block $i$, secrecy outage event occurs.
The secrecy outage event (with parameter $\epsilon$)
at block $i$ is defined as
\begin{align}
{\Oc}_{\text{sec}}(i,\epsilon) \triangleq {\Oc}_{\text{inf}}(i,\epsilon) \cup  {\Oc}_{\text{eq}}(i,\epsilon),  \label{eq:secrecyoutage}
\end{align}
where information outage occurs if accumulated mutual information on the message $W(i)$ remains below its entropy rate
\begin{align}
{\Oc}_{\text{inf}}(i, \epsilon) \triangleq\left\{ \frac{1}{N}I\left(W(i);Y^{iN} \right) < R_s - \epsilon  \right\},  \label{eq:informationoutage}
\end{align}
and the equivocation outage
occurs if the equivocation rate\footnote{Although the messages $\{W(i)\}_{i=1}^{M}$
are mutually independent, they may be
dependent conditioned on eavesdroppers' received signal $Z^{NM}$,
therefore equivocation expression includes conditioning on 
$W^{M} \backslash W(i)$.} of message $w(i)$ is less than $R_s -\epsilon$
\begin{align}
& \quad {\Oc}_{\text{eq}}(i, \epsilon) \triangleq \left\{ \frac{1}{N} H\big{(}W(i)|Z^{NM},
W^{M} \backslash W(i),g^{M}\big{)} < R_s -\epsilon \right\}. \label{eq:equivocationoutage}
\end{align}
\begin{definition}\label{secrecyoutageconstraint}\cite{gungor2013secrecy} 
Rate $R_s$ is achievable securely with at most $\alpha$ probability of secrecy outage if, 
for any fixed $\epsilon>0$, there exists a sequence of codes of rate no less than $R_s$ 
such that, for all large enough $N$, $M_1$ and $M_2$ such that $M=M_1M_2$, the conditions
\begin{align}
\Pr(E(i)|\bar{\Oc}_{\text{sec}}(i,\epsilon)) < \epsilon \label{errorconstraint}\\
\Pr ({\Oc}_{\text{sec}}(i,\epsilon)) < \alpha+\epsilon  \label{outage} 
\end{align}
are satisfied for all $i$ such that $i>M_1$, and for all possible adversary strategies $\phi^M$.
\end{definition}
The secrecy capacity with $\alpha$ outage is the supremum of such achievable secrecy rates.
We use $C_{s_d}(\alpha)$ to denote $\alpha$-outage secrecy capacity under no feedback, and use $C_{s_d}^{\text{1-bit}}(\alpha)$
to denote $\alpha$-outage secrecy capacity under 1-bit feedback at the end of each block.

Note that we do not impose a secrecy outage constraint on the first $M_1$ blocks, 
which is referred to as an initialization phase,
used to generate initial common randomness between the legitimate
nodes. Note that this phase only needs to appear \emph{once} in the communication lifetime of that link. 
In other words, when a session (which consists of $M$ blocks) between the associated nodes is over, they would have 
sufficient number of common key bits for the subsequent session,
and would not need to initiate the initialization step again \cite{gungor2013secrecy}.
\begin{theorem}(\textbf{Time sharing lower bound for $\alpha$-outage secrecy capacity})\label{t:delaynofeedback}
For no feedback,
$C_{s_d}(\alpha) \geq C_{s_d}^{-}(\alpha)$, where
\begin{align}
&\qquad C_{s_d}^{-}(\alpha)= \max_{\gamma,\tilde{R}_s,R_s} R_s \label{obj} \\
&\text{subject to: } \nonumber\\
&\Pr\bigg{(} \left\{(1-\gamma) \log\left(1+\frac{P_tH_m}{1+P_jH_z}\right)\geq \tilde{R}_s\right\} \bigcap 
\left\{ \left[\tilde{R}_s-(1-\gamma)\log(1+P_tH_e)\right]^{+}\geq R_s-R_{r0}  \right\}\bigg{)}\geq 1-\alpha \label{delay_nofb_outage}\\
&R_s\leq \tilde{R}_s, R_{r0}=\gamma C_s^{-},\gamma \in [0,1],\label{delay_nofb_keyrate}
\end{align}
where $C_s^{-}$ is provided in~\eqref{nofeedbackyx}.\qed
\end{theorem}
Similarly, for 1-bit feedback, $\alpha$-outage secrecy capacity is lower bounded by $C_{s_d}^{-\text{1-bit}}(\alpha)$, where $C_{s_d}^{-\text{1-bit}}(\alpha)$ is in the form (\ref{obj}-\ref{delay_nofb_keyrate}), except $R_{r0}$ is replaced with $R_{r1}=\gamma C_s^{-\text{1-bit}}$.

Here, we provide a sketch of achievability. The complete proof is in Appendix~\ref{app:delay_nofb}. In Theorem~\ref{t:delaynofeedback},  $\gamma \in [0,1]$ is the time-sharing parameter. We utilize the first $\gamma N$ channel uses
of each block to generate keys using the scheme described in proof of Theorem~\ref{thm:nocsi}. Using a code $(2^{NM_1R_{r0}},\gamma NM_1)$, we can generate
$NM_1R_{r0}$ secret key bits at the end of every $M_1$ blocks, where $R_{r0}\leq \gamma C_s^{-}$.
 These key bits are stored at the transmitter and the legitimate receiver, to help secure the delay sensitive messages
 in the following $M_1$ blocks.
We utilize the rest of the channel ($N(1-\gamma)$ channel use at each block) to send the delay constraint message. At each block $i$, $i>M_1$, message $w(i)$ of size $NR_s$ bits
is divided to two independent messages $w_1(i)$ and $w_2(i)$, of sizes $NR_{r0}$ and $N(R_s-R_{r0})$, respectively. Message $w_1(i)$ is secured 
via a one-time pad with the stored keys. Message $w_2(i)$ is secured with an additional randomization and the one-time padded message. With the following remark, we demonstrate the relation of the secrecy capacity with a delay constraint and the secrecy capacity without a delay constraint.
\begin{remark}(\textbf{Non-zero delay limited secrecy capacity})\label{nonzero_delay} Suppose that $H_m^{*}=\frac{P_tH_m}{1+P_jH_z}$ has a strictly monotone cdf and $\mathbb P(H_m^{*}\neq 0)>0$. If $\alpha\in (0,1]$ and $C_s^{-} > 0$, then $C_{s_d}(\alpha) > 0$. We can observe this fact by setting $\tilde{R}_s = R_s = R_{r0}$ in Theorem~\ref{t:delaynofeedback}.
Furthermore, note that $C_s^{\text{1-bit}}>0$ if $P(H_m^{*}\neq 0)>0$ (Remark~\ref{nonzero}). Hence, by setting $\tilde{R}_s = R_s = R_{r1}$, we can get $C_{s_d}^{\text{1-bit}}(\alpha) > 0$ for any $\alpha\in (0,1]$.

\qed
\end{remark}
\section{Numerical Evaluation}
\label{chap:numeric}
In this section, we conduct Monte Carlo simulations to illustrate our main results. We compare the secrecy capacity lower and upper bounds of the no feedback case with the lower bound of the secrecy capacity with 1-bit feedback.  To evaluate the effect of delay constraint, we also plot the lower bound of the $\alpha$-outage secrecy capacity with no feedback and 1-bit feedback. 
We consider that the power gains of the main, eavesdropper, and jamming channels independently follow an exponential distribution. 

\begin{figure}[h]
   \centering
   \includegraphics[width=0.5\textwidth]{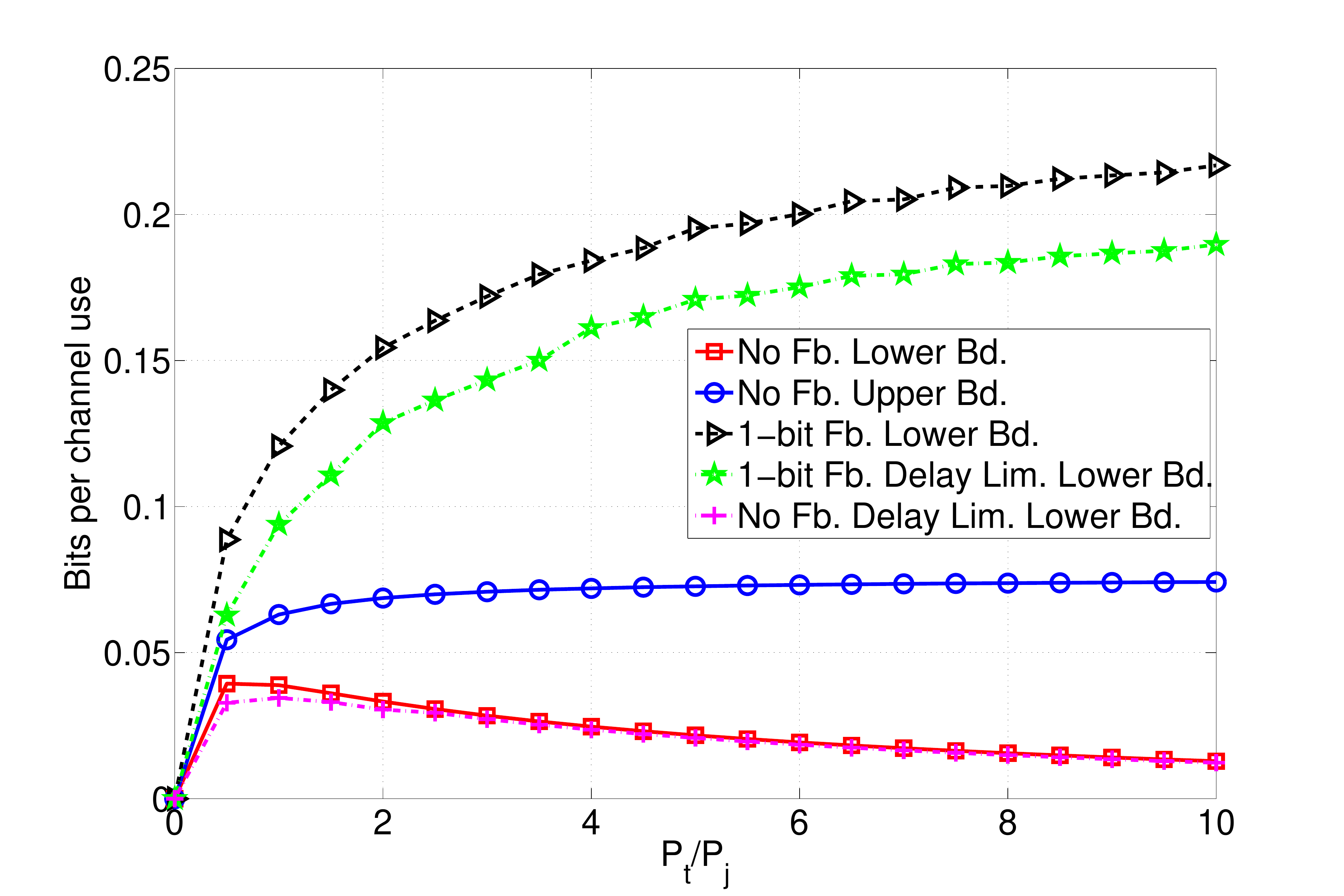}
   \caption{The comparison of the lower and upper bounds of the no feedback case with the lower bound of the 1-bit feedback case with  $\mathbb E[H_m]=5$, $\mathbb E[H_e]=2$, and $\mathbb E[H_z]=2$.}
   \label{fig:1}
 \end{figure}

In Figure~\ref{fig:1}, we fix the outage term $\alpha = 0.2$ and jamming power $P_j=1$, and we plot the secrecy capacity bounds as a function of the transmission power constraint, $P_t$. We take $\mathbb E[H_m]=5$, $\mathbb E[H_e]=2$, and $\mathbb E[H_z]=2$. A notable observation is that the lower bound for the no feedback case in Theorem \ref{thm:nocsi} decreases with $P_t$  beyond a certain point. The reason is that the lower bound, given in Theorem~\ref{thm:nocsi} is not always an increasing function of $P_t$ since the positive operator is outside of the expectation term. The lower bound to the $\alpha$-outage capacity without feedback, given in Theorem~\ref{t:delaynofeedback} also decreases with $P_t$ since the achievabilitiy strategy employs a key generation step in which keys are generated with the strategy used in the achievability proof of Theorem~\ref{thm:nocsi} . Let us replace $P_t$ in the lower bounds with dummy variable $P$. We conclude that the lower bounds in Theorems \ref{thm:nocsi} and \ref{t:delaynofeedback} can be further tightened by maximizing them over $P\in[0,P_t]$. From Figure~\ref{fig:1}, we observe that the secrecy capacity with 1-bit feedback is twice as large as  that with no feedback at $P_t/P_j=10$.

\begin{figure}[h]
   \centering
   \includegraphics[width=0.5\textwidth]{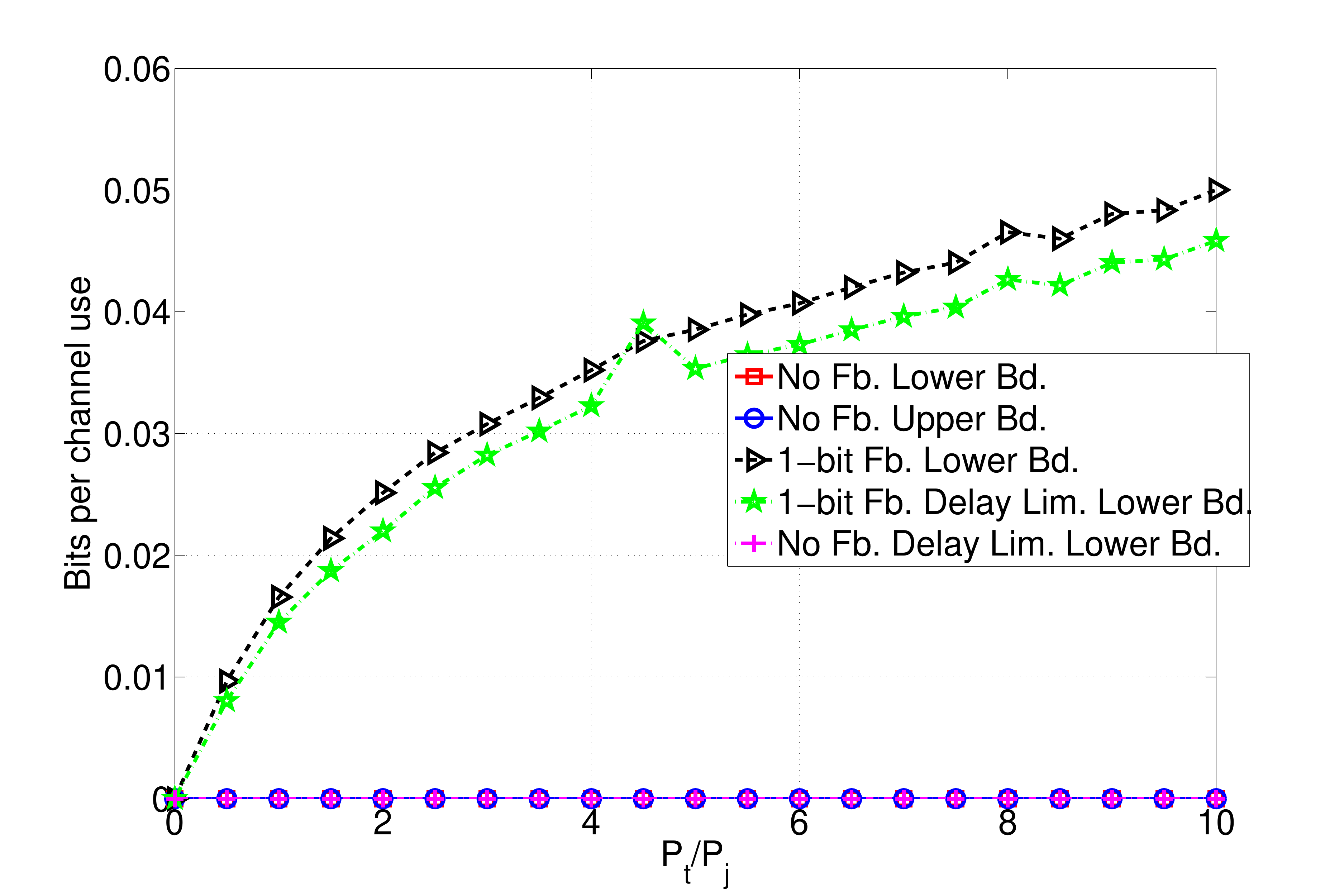}
    \caption{The comparison of the lower and upper bounds of the no feedback case with the lower bound of the 1-bit feedback case with  $\mathbb E[H_m]=1$, $\mathbb E[H_e]=2$, and $\mathbb E[H_z]=1$. }
\label{fig:2}
 \end{figure}

We now numerically illustrate Remark \ref{nonzero}, i.e., even when the eavesdropper channel is better \emph{on average}, we can achieve non-zero secrecy rates with the 1-bit feedback. We  take $\mathbb E[H_m]=1$, $\mathbb E[H_e]=2$, and $\mathbb E[H_z]=1$, i.e., the eavesdropper channel \emph{stochastically dominates} the effective main channel. As seen in Figure~\ref{fig:2}, we observe that  1-bit feedback sent at the end of each block is sufficient to make the secrecy capacity non-zero. Furthermore, we observe that the secrecy capacity of the no feedback case is zero (Remark~\ref{dom_remark}). The importance of the feedback can also be seen in the delay limited set-up, where no feedback strategy results in a zero achievable rate as opposed to the strategy employing 1-bit feedback.

\begin{figure}[h]
   \centering
   \includegraphics[width=0.5\textwidth]{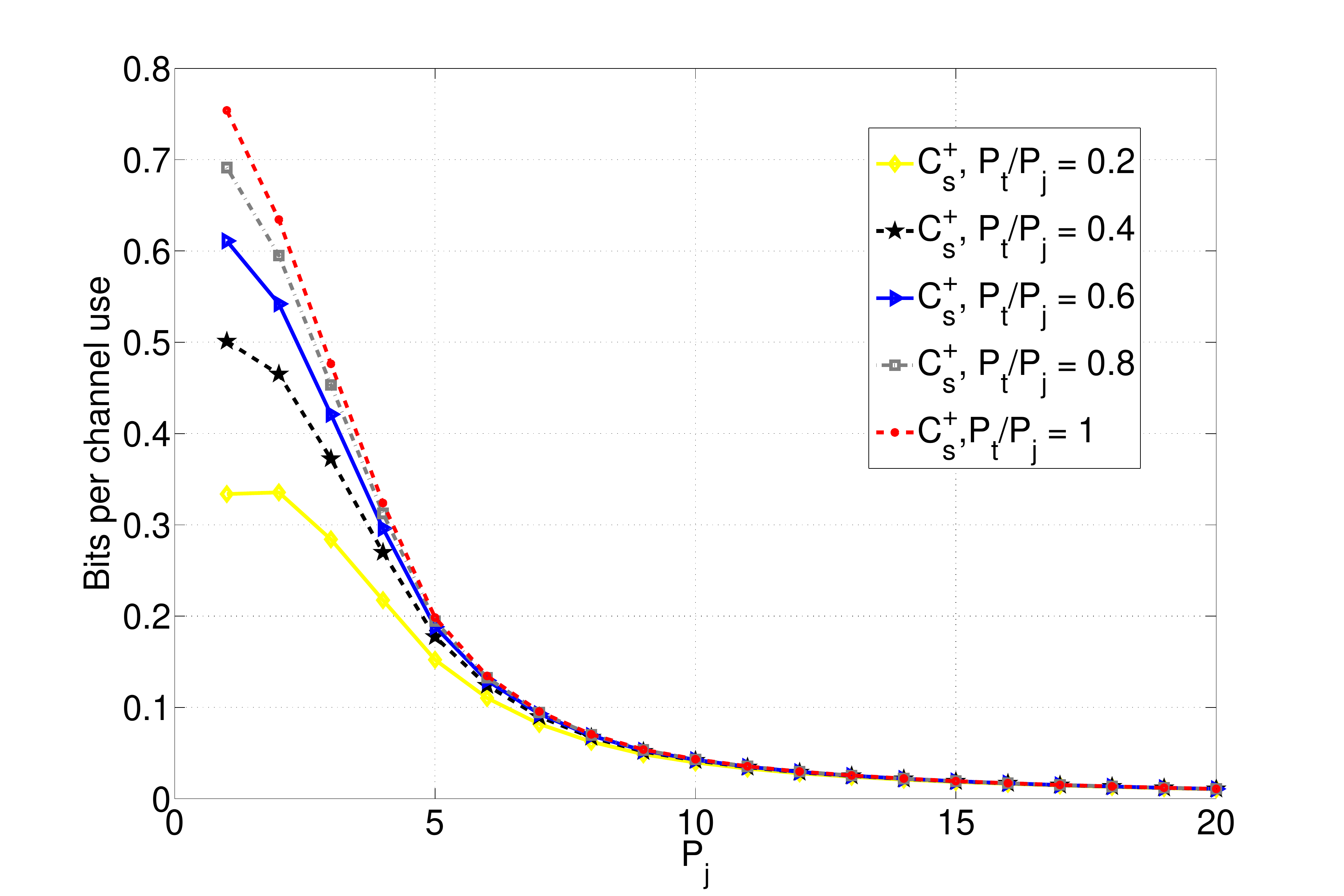}
    \caption{The change of the upper bound of the no feedback case when the transmission power constraint and jamming power scale in the same order.  $\mathbb E[H_m]=1$, $\mathbb E[H_e]=2$, and $\mathbb E[H_z]=1$. }
\label{fig:3}  
 \end{figure}

We illustrate Corollary~\ref{asym_power} in Figure~\ref{fig:3}.   For each plot in Figure~\ref{fig:3}, we keep the ratio of transmission power constraint and adversary power same, and we increase the jamming power. As mentioned in Corollary~\ref{asym_power},  in Figure~\ref{fig:3}, we observe  that the secrecy capacity with no feedback goes to zero, when the transmission power constraint and adversary power increase in the same order.


\section{Conclusion}
\label{chap:conc}
We study the impact of a hybrid adversary, that arbitrarily jams or eavesdrops at a given block, on the 
secrecy capacity of point to point Gaussian block fading channels.
We illustrate the necessity of receiver-to-transmitter feedback by considering two cases: 1) no feedback and 2) 1-bit feedback at the 
end of each block. For both cases, we bound the secrecy capacities. We show that, without any feedback, the secrecy capacity is zero if
the eavesdropper channel power gain stochastically dominates the effective main channel power gain.
We also observe that, the secrecy capacity vanishes asymptotically when 
the transmit power constraint and jamming power increase in the same order. 
However, even with 1-bit receiver feedback at the end of each block, the secrecy
capacity is non-zero for the wide class of channel statistics  as described in Remark \ref{nonzero}.
We also analyze the effects of multiple colluding/non-colluding adversaries and delay.
We show that,  with no feedback, multiple adversaries can hurt the secrecy capacity
even more, as the secrecy capacity bounds are not affected by the cross-interference across the adversaries.
Finally, we provide a novel time-sharing approach for the delay limited setting, and we show that  $\alpha$-outage secrecy capacity
is positive whenever the secrecy capacity without any delay limitation is positive (Remark~\ref{nonzero_delay}).
\appendices
\section{Proof of Theorem~\ref{thm:nocsi}} \label{section:nocsi}
\emph{Codebook Generation:} Pick  $R_s= C_s^{-}$ and $R_m=\mathbb E\left[\log\left(1+\frac{P_tH_m}{1+P_jH_z}\right)\right]-\epsilon$ for some
 $\epsilon > 0$. Generate codebook $\mathit{c}$ containing independently and identically generated codewords $x^{NM}_l, l\in[1:2^{NMR_m}]$, 
each of which are	 drawn from $\prod_{k=1}^{NM} p_X(x_{lk})$. Here, $p_X(x)$ is the probability density function of complex Gaussian random variable with zero mean and variance $P_t$.

\emph{Encoding:} To send message $w \in [1:2^{NMR_s}]$, the secrecy encoder draws index $l$ from the uniform distribution whose sample 
space is $\left[(w-1)2^{NM(R_m-R_s)}+1:w2^{NM(R_m-R_s)}\right]$. The channel encoder then transmits corresponding codeword, $x^{NM}_l$.

\emph{Decoding:} Let $y^{NM}$ be the received sequence. If the adversary is in the eavesdropping state, i.e, $\phi(i)=0$, the channel decoder
 draws $g_z(i)$ from $G_z(i)$ and a noise sequence $s_j^N(i)$ from $S_j^N(i)$ to obtain
 $$\hat{y}^N(i) =y^{N}(i)+g_z(i)s_j^N(i). $$The channel decoder looks for a unique message $w\in [1:2^{NMR_s}]$ such that 
$\left(x_l^{NM}, (\hat{y}^{NM},g_m^M, g_z^M)\right)\in A^{NM}_{\epsilon}$, where 
$A^{NM}_{\epsilon}\left(X^{N}, (\hat{Y}^{N},G_m, G_z)\right)$ is the set 
of jointly typical $\left(x^{NM}, (\hat{y}^{NM}, g_m^M, g_e^M)\right)$ sequences with 
\begin{equation}
\hat{Y}^N = G_mX^N+ G_zS_j^N + S_m^N
\end{equation} 

\emph{Analysis of the probability error and secrecy:} Random coding argument is used to show that there exists
 sequences of codebooks that satisfy the constraint (\ref{cond1}) and (\ref{cond2}) simultaneously. Since 
$R_m < \frac{1}{N} I\left(X^N; \hat{Y}^N,G_m^M, G_z^M\right)= \mathbb E\left[\log\left(1+\frac{P_tH_m}{1+P_jH_z}\right)\right]$, by the channel coding 
theorem~\cite{cover1991}, we have $\mathbb E_{\mathcal{C}}(P_{\epsilon}^{NM}(\mathcal{C})) \to 0 \text{ as } M \to \infty$, where the expectation is over all random codebooks. We show below that $\mathbb E_{\mathcal{C}}\left[R_e(\mathcal{C})\right] \to R_s \text{ as } M \to \infty$, where $$\mathbb E_{\mathcal{C}}\left[R_e(\mathcal{C})\right]=H(W|Z^{NM}, g^M, 
\mathcal{C}).
$$ Hence, there exists a sequences of codebooks that satisfy both (\ref{cond1}) and (\ref{cond2}) since we have
 $\mathbb E_{\mathcal{C}}\left[P_{\epsilon}^{NM}(\mathcal{C})+R_e(\mathcal{C})\right] \to R_s$ as $M \to 0$.

For the secrecy analysis,  let's define $\hat{Z}^N(i) =X^{N}(i) g_e(i)+S_e^{N}(i)$, $1 \leq \forall i \leq M$. 
The equivocation analysis averaged over codebooks is as follows. \allowdisplaybreaks
\begin{align}
&MNR_e(\mathcal{C})=H(W|Z^{NM}, g^M, \mathcal{C}) \nonumber \\
&=H(W|Z^{NM}, h_e^M,\mathcal{C})\nonumber\\
&\stackrel{(a)}{\geq} H(W|\hat{Z}^{NM}, g_e^M,\mathcal{C}) \nonumber \\
&= H(W, X^{NM}|\hat{Z}^{NM},h_e^M,\mathcal{C})-H(X^{NM}|\hat{Z}^{NM}, W, g_e^M,\mathcal{C})\nonumber\\
&  = H(X^{NM}|\hat{Z}^{NM}, h_e^M,\mathcal{C})+H(W|X^{NM},\hat{Z}^{NM}, g_e^M,\mathcal{C})\nonumber\\
&\qquad\qquad\qquad\qquad\qquad\qquad-H(X^{NM}|\hat{Z}^{NM}, W, g_e^M,\mathcal{C})\nonumber\\
&\geq H(X^{NM}|\hat{Z}^{NM}, h_e^M,\mathcal{C})+H(X^{NM}|\hat{Z}^{NM}, W, g_e^M,\mathcal{C})\nonumber\\
&= H(X^{NM}|h_e^{M})-I(X^{NM}; \hat{Z}^{NM}|h_e^M,\mathcal{C})\nonumber\\
&\qquad\qquad\qquad\qquad\qquad\qquad+H(X^{NM}|\hat{Z}^{NM}, W, g_e^M, \mathcal{C})\nonumber\\
&\stackrel{(b)}{=}MN R_m - I(X^{NM}; \hat{Z}^{NM}|g_e^M,\mathcal{C})\nonumber \\
&\qquad\qquad\qquad\qquad\qquad\qquad-H(X^{NM}|\hat{Z}^{NM}, W, g_e^M,\mathcal{C})\nonumber\\
&\geq MN R_m - I(X^{NM},\mathcal{C}; \hat{Z}^{NM}|g_e^M,)\nonumber \\
&\qquad\qquad\qquad\qquad\qquad\qquad-H(X^{NM}|\hat{Z}^{NM}, W, g_e^M,\mathcal{C})\nonumber\\
&\stackrel{(c)}{=}MN R_m - I(X^{NM}; \hat{Z}^{NM}|g_e^M)\nonumber \\
&\qquad\qquad\qquad\qquad\qquad\qquad-H(X^{NM}|\hat{Z}^{NM}, W, g_e^M,\mathcal{C})\nonumber\\
&\geq MN R_m - N\sum_{i=1}^M \log(1+P_th_e(i))\nonumber\\
&\qquad\qquad\qquad\qquad\qquad\qquad -H(X^{NM}|\hat{Z}^{NM}, W, g_e^M,\mathcal{C}) \nonumber
\end{align}
where $(a)$ follows from the fact that $W\to Z^{NM}, G_e^M\to G_m^M,G_z^M$ forms a Markov chain, $(b)$ follows from the fact that conditioning reduces the entropy, $(c)$ follows from the fact that codeword 
$X^{NM}$ is uniformly distributed over a set of size $2^{NMR_m}$, and $(d)$ follows from the fact that 
\begin{equation} 
\mathcal{C} \to X^{NM} \to \hat{Z}^{NM} 
\end{equation}
forms Markov chain. We continue with the following steps.
\begin{align}
&\frac{1}{MN}H\left(W|Z^{NM}, g_e^M,\mathcal{C}\right) \nonumber\\
&\geq R_m-\sum_{i=1}^M \frac{1}{M}\log(1+P_th_e(i))\nonumber\\
&\qquad\qquad\qquad-\frac{1}{MN}H\left(X^{NM}|\hat{Z}^{NM}, W, g_e^M,\mathcal{C}\right)\nonumber\\
&\stackrel{(e)}{\geq}R_m- \mathbb E\left[\log(1+P_tH_e)\right]-\epsilon_1\nonumber \\
&\qquad\qquad\qquad\qquad-\frac{1}{MN}H\left(X^{NM}|\hat{Z}^{NM}, W, g_e^M,\mathcal{C}\right) \nonumber\\
&\stackrel{(f)}{\geq}R_m- \mathbb E\left[\log(1+P_tH_e)\right]-\epsilon_1-\epsilon_2\nonumber\\
&=R_s -\epsilon-\epsilon_3,\nonumber
\end{align}
where $\epsilon_3 = \epsilon_1+\epsilon_2$. Here, for any $\epsilon_1>0$, $(e)$ is satisfied  for any $h_e^M\in B_M$ with $Pr[B_M]=1$ and for sufficiently large $M$ since $$\lim_{M\rightarrow \infty}\frac{1}{M}\sum_{i=1}^M\log(1+P_tH_e(i))=\mathbb E\left[\log(1+P_tH_e)\right]$$\text{with probability 1}, and $(f)$ follows from the Fano's inequality. Let's define $R_{me}\triangleq R_m-R_s$ and $E^{NM} \triangleq \mathbb E_{\mathcal{C}}\left[P\left(X^{NM} \neq \hat{X}^{NM}|W=w,h_e^M,\mathcal{C}\right)\right]$, where $\hat{X}^{NM}=g(\hat{Z}^{NM},g_e^M, W=w, \mathcal{C}=c)$ is the estimation of the codeword $X^{NM}$.
\begin{align}
&\frac{1}{MN}H\left(X^{NM}|\hat{Z}^{NM}, W=w, g_e^M,
\mathcal{C}\right) \leq E^{NM} R_{me} +\frac{1}{MN}H(E^{NM})\label{fano1}\\
&\qquad\qquad\qquad\qquad\qquad\qquad\qquad\qquad                 \leq \epsilon_2 \label{fano}
\end{align}
Here, for any $\epsilon_2>0$ and $w\in[1:2^{NMR_s}]$, (\ref{fano}) is satisfied for sufficiently large $M$. The reason is that since $R_{me}=I(X^{N},\hat{Z}^{N}|H_e)-\epsilon$, $E^{NM} \to 0$ as $M \to \infty$ from the random coding argument~\cite{cover1991}.\qed

We now provide the proof of the upper bound in Theorem~\ref{thm:nocsi}. Suppose that $R_s$ is achievable rate. From definition~(\ref{cond1})-(\ref{cond2}) and Fano's inequality, we have
\begin{align}
&\min_{\phi(i):1\leq i \leq M} \frac{1}{NM} H\left(W|Z^{NM},g^M,\phi^M\right) \geq R_s-a_{NM} \label{convMain1}\\
&\max_{\phi(i):1\leq i \leq M} \frac{1}{NM} H\left(W|Y^{NM},g^M,\phi^M\right)\leq b_{NM} \label{convEave1}
\end{align}
for any $ h^M\in \mathcal{A}_M$ with $\mathbb P(\mathcal{A}_M) \geq 1-c_{NM}$.Here, $a_{NM}$, $b_{NM}$, and $c_{NM}$ go to zero as $N\to \infty$ and $M\to \infty$. 

Adversary strategy $\phi(i)=0$, $1\leq \forall i \leq M$ solves LHS of (\ref{convMain1}) and strategy $\phi(i)=1$, $1\leq \forall i \leq M$ solves LHS of (\ref{convEave1}).
Hence, we have 
\begin{align}
& \frac{1}{NM} H\left(W|\hat{Z}^{NM},g^M\right) \geq R_s-a_{NM} \label{convMain2}\\
& \frac{1}{NM} H\left(W|\hat{Y}^{NM},g^M\right)\leq b_{NM} \label{convEave2}
\end{align}
where
\begin{align}
&\hat{Y}^N(i) = g_m(i)X^N(i)+ g_z(i)S_j^N(i)+ S_m^N(i), \text{ and }\\ &\hat{Z}^N(i)=g_e(i)X^N(i) + S_e^N(i),  \qquad 1\leq \forall i \leq M.
\end{align}
Here, the LHS of \eqref{convMain1} equals to that of \eqref{convMain2} since conditioning reduces the entropy and the LHS of \eqref{convEave1} equals to that of \eqref{convEave2} since $W\to Y^{NM}\to \hat{Y}^{NM}$ forms a Markov chain.

We now show that if $R_s$ is achievable, we have 
\allowdisplaybreaks
\begin{align}
& \frac{1}{NM} H(W| \hat{Z}^{NM}, H^M)\geq \label{llll1} \\
&\int_{ \mathcal{A}_M}\frac{1}{NM} H(W| \hat{Z}^{NM},  g^M) f_{H^M} ( h^M) \;dh^M\nonumber\\
&\geq \int_{\mathcal{A}_M}(R_s -a_{NM}) f_{G^M} ( g^M) \;dg^M\label{sketch_def}\\
&\geq R_s-\delta_{NM}, \label{sketch_def2}
\end{align}
where $G^M=\left[G_m^M, G_e^M, G_z^M\right]$, $\delta_{NM} = -R_s c_{NM}-a_{NM}+a_{NM}c_{NM}$, and $\delta_{NM}\to 0$ as $N\to \infty$ and $M\to \infty$. Here, (\ref{sketch_def}) follows from \eqref{convMain2}, and (\ref{sketch_def2}) follows from the fact that $P[\mathcal{A}_M]\geq1-c_{NM}$. Note that here, the message $W$ is conditioned on random vector, $G^M$ instead of $g^M$ in \eqref{convMain2}. With the similar steps, we can show that 
\begin{align}
\frac{1}{NM} H(W| \hat{Y}^{NM}, G^M) \leq \epsilon_{NM}, \label{sketch_def3}
\end{align}
where $\epsilon_{NM}\to 0$ as $N\to \infty$ and $M\to \infty$.
The upper bound, $C_s^{+}$ follows when we combine \eqref{sketch_def2} and \eqref{sketch_def3} with the following steps: 
\begin{align}
&R_s \leq \frac{1}{NM} H(W| \hat{Z}^{NM}, G^M)\nonumber\\ 
&\qquad-\frac{1}{NM} H(W| \hat{Y}^{NM}, G^M)+\gamma_{NM}\\
&\stackrel{(a)}{=}\frac{1}{NM} H(W| \tilde{Z}^{NM}, \tilde{G}_m^M,\tilde{G}_e^M,\tilde{G}_z^M)\nonumber\\
&\qquad\qquad-\frac{1}{NM} H(W| \tilde{Y}^{NM}, \tilde{G}_m^M,\tilde{G}_e^M,\tilde{G}_z^M)+\gamma_{NM} \label{newRV}\\
&=\frac{1}{NM}I(W; \tilde{Y}^{NM}|\tilde{Z}^{NM},\tilde{G}_m^M,\tilde{G}_e^M,\tilde{G}_z^M)+\gamma_{MN}\\
&\stackrel{(b)}{\leq} \frac{1}{NM}I(X^{NM}; \tilde{Y}^{NM}|\tilde{Z}^{NM},\tilde{G}_m^M,\tilde{G}_e^M,\tilde{G}_z^M)+\gamma_{MN}\\
&\stackrel{(c)}{\leq}\frac{1}{NM}\sum_{i=1}^{M} I(\tilde{X}^N(i),\tilde{Y}^N(i)|\tilde{Z}^N(i),\tilde{G}_m(i),\tilde{G}_e(i),\tilde{G}_z(i))+ \gamma_{NM}\\ 
&\stackrel{(d)}{\leq}\frac{1}{NM}\sum_{i=1}^{M}\sum_{j=1}^N I\bigg{(}X(i,j),\tilde{Y}(i,j)|\tilde{Z}(i,j),\tilde{G}_m(i), \tilde{G}_e(i),\tilde{G}_z(i)\bigg{)}+ \gamma_{NM}\\ 
&\stackrel{(e)}{\leq} \frac{1}{NM}\sum_{i=1}^{M}\sum_{j=1}^N \mathbb E\left[\left(\log\left(1+\frac{ P_{t_{ij}}\tilde{H}_m}{1+P_j\tilde{H}_z}\right)-\log\left(1+ P_{t_{ij}}\tilde{H}_e\right)\right)^{+}\right]  + \gamma_{NM}\\
&\stackrel{(f)}{\leq} \mathbb E\left[\left(\log\left(1+\frac{\frac{1}{NM}\sum_{i=1}^{M}\sum_{j=1}^N P_{t_{ij}}\tilde{H}_m}{1+P_j\tilde{H}_z}\right)-\log\left(1+\frac{1}{NM}\sum_{i=1}^{M}\sum_{j=1}^N P_{t_{ij}}\tilde{H}_e\right)\right)^{+}\right]+ \gamma_{NM}\label{jensen_1}\\
&\stackrel{(g)}{\leq}  \mathbb E\left[\left(\log\left(1+\frac{P_t\tilde{H}_m}{1+P_j\tilde{H}_z}\right)-\log\left(1+P_t\tilde{H}_e\right)\right)^{+}\right]+\gamma_{NM},
\end{align}
where the notation $(i,j)$ indicates the $j$-th channel use of $i$-th block and $\gamma_{NM}=\delta_{NM}+\epsilon_{NM}$. Note that $\gamma_{NM}\to 0$ as $N$ and $M \to \infty$. In~
\eqref{newRV}, we define new random variables, i.e.,

\begin{align}
&\tilde{Y}^N(i) = \tilde{G}_m(i)X^N(i)+ \tilde{G}_z(i)S_j^N(i)+ S_m^N(i), \text{ and }\\ 
&\tilde{Z}^N(i)=\tilde{G}_e(i)X^N(i) + S_e^N(i),  \qquad 1\leq \forall i \leq M.
\end{align}
and $\tilde{H}_m(i)=|\tilde{G}_m(i)|^{2}$, $\tilde{H}_e(i)=|\tilde{G}_e(i)|^{2}$, and $\tilde{H}_z(i)=|\tilde{G}_z(i)|^{2}$. Here, $\left\{\tilde{G}_e(1), \tilde{G}_e(2),\dots, \tilde{G}_e(M)\right\}$ are i.i.d random variables with $\tilde{G}_e(i) \sim p_{G_e}$, and $G_e^M$ is independent from $\left( W, S_e^{NM},S_j^{NM}, S_m^{NM}\right)$. In a similar way, $\left\{\left(\tilde{G}_m(1),\tilde{G}_z(1)\right), \left(\tilde{G}_m(2),\tilde{G}_z(2)\right),\dots, \left(\tilde{G}_m(M),\tilde{G}_z(M)\right)\right\}$ are i.i.d random vectors with $\left(\tilde{G}_m(i),\tilde{G}_z(i)\right) \sim p_{G_m, G_z}$, and $\left(G_m^M, G_z^M\right)$ are independent from $\left( W, S_e^{NM},S_j^{NM}, S_m^{NM}\right)$.
\\

For the derivation above, $(a)$ follows from the fact $\left(W, Z^{NM}, G_e^{M}\right)$ and $\left(W, Y^{NM}, G_e^{M}\right)$ have the same joint pdf with $\left(W, \tilde{Z}^{NM}, \tilde{G}_e^{M}\right)$ and $\left(W, \tilde{Y}^{NM}, \tilde{G}_e^{M}\right)$, respectively. Furthermore, note that $W\to \hat{Z}^{NM},G_e^M\to G_m^M,G_z^M $ and $W\to \tilde{Z}^{NM},\tilde{G}_e^M\to \tilde{G}_m^M,\tilde{G}_z^M $ form Markov chain. In a similar way, $W\to \hat{Y}^{NM},G_m^M,G_z^M \to G_e^M$ and $W\to \tilde{Y}^{NM}, \tilde{G}_m^M,\tilde{G}_z^M \to \tilde{G}_e^M$ form Markov chain.
$(b)$ follows from the fact that $W\to X^{NM},\tilde{Z}^{NM},\tilde{G}_m^M, \tilde{G}_e^M, \tilde{G}_z^M\to \tilde{Y}^{NM}$ forms a Markov chain.  $(c)$ and $(d)$ follows from the memoryless property of the channel and from the fact conditioning reduces the entropy.

The power constraint in~\eqref{pw_cst} implies that  $\frac{1}{NM}\sum_{i=1}^{M}\sum_{j=1}^N \mathbb E\left[|X(i,j)|^2\right]\leq P_t$, where the expectation is taken over $W$. Also, note that 
$\tilde{G}(i)=\left[\tilde{G}_m(i), \tilde{G}_e(i),\tilde{G}_z(i)\right]$ and $X(i,j)$ are independent random variables. Define $P_{{t}_{ij}}\triangleq\mathbb E\left[|X(i,j)|^2\right]= \mathbb E\left[|X(i,j)|^2|\tilde{G}(i)=g(i)\right]$. Then, $(e)$ follows from the fact that Gaussian distribution maximizes the conditional mutual information~\cite{leung1978}. In~\eqref{jensen_1}, $(f)$ follows from the fact that $\left(\log(1+P_{t_{ij}}x)-\log(1+P_{t_{ij}}y)\right)^{+}$ is a concave function of $P_{t_{ij}}$ for any $x\geq 0$ and $y\geq0$ and from Jensen's inequality. Finally, $(g)$ follows from the fact that $\left(\log(1+Px)-\log(1+Py)\right)^{+}$ is a non-decreasing function in $P$ for any $x\geq 0$ and $y\geq0$.
.\qed
\section{Proof of Corollary \ref{asym_power}}
\label{high_power_proof}
We have the following analysis:
\begin{align}
&\lim_{P\to\infty} C_s^{+}\\
&\hspace{-0.5cm}\leq\lim_{P\to \infty}\mathbb E\left[\left(\log\left(1+\frac{P_t(P)H_m}{1+P_j(P)H_z}\right)-\log\left(1+P_t(P)H_e\right)\right)^{+}\right]\nonumber\\
&\hspace{-0.5cm}\stackrel{(a)}{=}\mathbb E\left[\lim_{P\to \infty}\left(\log\left(1+\frac{P_t(P)H_m}{1+P_j(P)H_z}\right)-\log\left(1+P_t(P)H_e\right)\right)^{+}\right]\nonumber\\
&=0.
\end{align}
Here, $(a)$ follows from the dominant convergence theorem. To apply dominant convergence theorem,  we need to show that 
\begin{align} 
&g_P(H_m,H_e,H_z)=\left(\log\left(1+\frac{P_t(P)H_m}{1+P_j(P)H_z}\right)-\log\left(1+P_t(P)H_e\right)\right)^{+}\label{objective}
\end{align}
is upper and lower bounded by random variables that have a finite expectation. Note that $g_P(H_m,H_e,H_z)$ is  lower bounded by zero and upper bounded by $\log\left(1+ \frac{P_t(P)H_m}{P_j(P)H_z}\right)$ with probability 1.

Since $P_t\left(P\right)= \mathcal{O}\left(P_j(P)\right) \text{ as } P\to\infty$, there exists finite $B$ and $p_0$ such that $P_t(P) \leq B \times P_j(P)$ for all $P>p_0$.
 We now show that $\mathbb E[g_P(H_m,H_e,H_z)]$ has a finite expectation for all $P>p_o$ with the following analysis:
\begin{align}
&\mathbb E[g_P(H_m,H_e,H_z)]\leq \mathbb E\left[\log\left(1+ \frac{B H_m}{H_z}\right)\right]\\
&= \mathbb E[\log(H_z+BH_m)] -\mathbb E[\log(H_z)] \nonumber \\
&\leq \log\left(B\mathbb E[H_m]+\mathbb E[H_z]\right)-\mathbb E[\log(H_z)] \label{jensen}\\
&\leq \log\left(B\mathbb E[H_m]+\mathbb E[H_z]\right) -\int_0^1 \log(h_z) f_{H_z}(h_z) \;d h_z\\
&\leq \log\left(B\mathbb E[H_m]+\mathbb E[H_z]\right) -A\int_0^1 \log(h_z) \;d h_z\\
&=\log\left(B\mathbb E[H_m]+\mathbb E[H_z]\right) +A\log(e) \label{log}\\
& < \infty, \label{bound}
\end{align}
for all $P> p_0$, where $A = \sup_{h_z} f_{H_z}(h_z)$. Here, (\ref{jensen}) follows from the Jensen's inequality, (\ref{log}) follows from the fact that $\int_0^1 \log(h_z)=-\log(e)$,  and (\ref{bound}) follows from the fact that $\mathbb E[H_m], \mathbb E[H_z] < \infty$ and the pdf of $H_z$  is bounded.

Since  $\log\left(1+ \frac{P_t(P)H_m}{P_j(P)H_z}\right)$ is a continuous function of $P$, it is a bounded function on the closed interval $[0, p_0]$ with probability 1. Hence, $\mathbb E[g_P(H_m,H_e,H_z)]< \infty$ for all $P\geq0$.\qed

\section{Proof of Theorem \ref{multi_no_csi}}
\label{multi_no_csi_proof}
The decoding and encoding strategies are the same with the strategies used in the proof of Theorem~\ref{thm:nocsi}. Therefore, we omit the probability error analysis and only focus on the secrecy analysis. We pick $R_m=\mathbb E\left[\log\left(1+\frac{P_tH_m}{1+P_j\hat{H}_z}\right)\right]-\epsilon$ for some $\epsilon >0$.

For the secrecy analysis, let's define 
$\hat{Z}_s^N(i) =X^{N}(i) g_{e_s}(i)+S_{e_s}^{N}(i)$, $1 \leq \forall s \leq S,\;1 \leq \forall i \leq M$.  
With the same steps used in the secrecy analysis of the proof of Theorem~\ref{thm:nocsi}, we can get
\begin{align}
&\frac{1}{MN} H\left(W\big\vert \left\{Z_s^N(i), g_{e_s}(i), \phi_s(i)\right\}_{ 1\leq i \leq M},\mathcal{C}\right)\\
&\geq L_s-\epsilon_1-\frac{1}{MN}H\left(X^{NM}|\hat{Z}_s^{NM}, W, g_{e_s}^M,\mathcal{C}\right)\label{m1234}
\end{align}
for any $\epsilon_1 > 0$ and sufficiently large $M$, where 
$$
L_s = \mathbb E\left[\log\left(1+\frac{P_tH_m}{1+P_j\hat{H}_z}\right)-\log\left(1+P_tH_{e_s}\right)\right].
$$
We now show that, for any $\epsilon_2>0$,
\begin{align}
&\frac{1}{MN}H\left(X^{NM}|\hat{Z}_s^{NM}, W, g_e^M,\mathcal{C}\right) \leq L_s -C_s^{NC-}+\epsilon_2\label{m12}
\end{align}
for sufficiently large $M$. To prove~(\ref{m12}), suppose that codewords correspond to message $W$ is partitioned into  $2^{NM(L_s-C_s^{NC-})}$ groups. Let's define random variable $T$ that represents the group index of $X^{NM}$. Then, we have 
\begin{align}
&\frac{1}{MN}H\left(X^{NM}|\hat{Z}_s^{NM}, W, g_{e_s}^M,\mathcal{C}\right) \label{m133}\\
&\leq\frac{1}{MN}H\left(X^{NM}, T|\hat{Z}_s^{NM}, W, g_{e_s}^M,\mathcal{C}\right)\\
&\hspace{-0.1cm}\leq\frac{1}{MN}H\left(X^{NM}|T,\hat{Z}_s^{NM}, W, g_{e_s}^M,\mathcal{C}\right)+\frac{1}{MN}H(T)\\
&\leq \epsilon_2 + L_s -C_s^{NC-}, \label{m13}
\end{align}
for any $\epsilon_2>0$ and sufficiently large $M$. Here, (\ref{m13}) follows from the random coding argument as in (\ref{fano1})-(\ref{fano}) of the proof of Theorem~\ref{thm:nocsi}. The proof follows when we combine (\ref{m1234}) and (\ref{m12}).\qed

We now provide the upper bound. Suppose that $R_s$ is achievable rate. From definition~(\ref{cond1})-(\ref{cond2}) and Fano's inequality, we have
\begin{align}
&\min_{1\leq s\leq S}\min_{\substack{\phi^M_j\\ 1\leq j\leq S}} \frac{1}{NM} H\left(W|Z_s^{NM},g^M,\{\phi^M_j\}_{ 1\leq j \leq S}\right) \nonumber\\
&\qquad\qquad\qquad\qquad\qquad\qquad\qquad\qquad \geq R_s-a_{NM} \label{convMain11}\\
&\max_{\substack{\phi^M_j\\ 1\leq j\leq S}} \frac{1}{NM} H\left(W|Y^{NM},g^M,\{\phi^M_j\}_{ 1\leq j \leq S}\right)\nonumber\\
&\qquad\qquad\qquad\qquad\qquad\qquad\qquad\qquad \leq b_{NM} \label{convEave11}
\end{align}
for any $g^M\in \mathcal{A}_M$ where $g^M  =\left[g_m^M, g_{e_1}^M , \dots g_{e_S}^M,  g_{z_1}^M, \dots g_{z_S}^M, \{g^M_{f_{sj}}\}_{1\leq s,j\leq S}  \right]$ with $\mathbb P(\mathcal{A}_M) \geq1-c_{NM}$. Here, $a_{NM}, b_{NM}$, and $c_{NM}$ go to zero as $N\to \infty$ and $M\to \infty$. 

For each adversary $s$, the adversary strategy $\phi_j(i)=0$, $1\leq \forall i \leq M,1\leq \forall j \leq S$ solves the inner minimization problem in the LHS of (\ref{convMain11}). The strategy $\phi_j(i)=1$, $1\leq \forall i \leq M,1\leq \forall j \leq S$ solves LHS of (\ref{convEave11}).
Hence, we have 
\begin{align}
&  \min_{1\leq s \leq S}\frac{1}{NM}H\left(W|\hat{Z}_s^{NM},g^M\right) \geq R_s-a_{NM} \label{convMain22}\\
& \frac{1}{NM} H\left(W|\hat{Y}^{NM},g^M\right)\leq b_{NM} \label{convEave22}
\end{align}
where
\begin{align}
&\hat{Y}^N(i) = g_m(i)X^N(i)+ \sum_{s=1}^S g_{z_s}(i)S_{j_s}^N(i)+ S_m^N(i)\\ 
&\hat{Z}_s^N(i)=g_{e_s}(i)X^N(i) + S_e^N(i)
\end{align}
 for $1\leq \forall i \leq M$ and  $1\leq \forall s \leq S$.
Here, the LHS of \eqref{convMain11} equals to that of \eqref{convMain22} since $W\to \hat{Z}_s^{NM}\to Z_s^{NM} $  and the LHS of \eqref{convEave11} equals to that of \eqref{convEave22} since $W\to Y^{NM}\to \hat{Y}^{NM}$ forms a Markov chain. Furthermore, note that $W\to \hat{Y}^{NM}, G^M_m, \{G_{z_s}\}_{1\leq s\leq S} \to G^M\backslash G^M_m, \{G_{z_s}\}_{1\leq s\leq S}$  and $W\to \hat{Z}^{NM}, \{G_{e_s}\}_{1\leq s\leq S} \to G^M\backslash \{G_{e_s}\}_{1\leq s\leq S}$ form Markov chains. The rest of the proof is similar to the proof of the upper bound given in Theorem~\ref{thm:nocsi}.\qed
\section{Proof of Theorem~\ref{1bitrate_accum}}
\label{app:ach_csi}
\emph{Codebook Generation:} Fix $R>0$ and $\epsilon >0$.  Pick $R_m = \frac{R}{\mathbb E[T]}-\epsilon$, where $T$ is defined in Theorem~\ref{1bitrate_accum}. Generate codebook $\mathit{c}$ containing independently and identically generated codewords $x_l^{N}, l\in[1:2^{NR}]$, each are drawn from $\prod_{k=1}^{N} p_X(x_{lk})$. Here, $p_X(x)$ is the distribution of complex Gaussian random variable with zero mean and variance $P_t$.
\\

\emph{Encoding:}  Pick M such that 
$\lvert\frac{1}{MN} R^{*}_w(M)- \frac{R}{\mathbb E[K]}\rvert \leq \epsilon$ with probability 1. Here, $R^{*}_w(M)$ is the accumulated reward at the receiver up to $M$-th block for the renewal process explained in the proof sketch of Theorem~\ref{1bitrate_accum}, where the reward at each renewal is $NR$ bits. To send a message $w\in [1:2^{NMR_s}]$, the secrecy encoder draws an index $l$ from the uniform distribution whose sample space is$ \left[(w-1)2^{NM(R_m-R_s^{\text{1-bit}})}+1:2^{NM(R_m-R_s^{\text{1-bit}})}\right]$.  Then, the secrecy encoder maps $l$ into $NMR_m$ bits and decompose $NMR_m$ bits into groups of $NR$ bits. To send the index $l$, the channel encoder transmits $NR$ in each block by using codebook  $\mathit{c}$. When NAK is received, the channel encoder sends the same bit group transmitted at the previous block. Detailed information about the encoding can be found in the proof sketch of Theorem~\ref{1bitrate_accum}.\\

\emph{Decoding:} Let $y^{N}(i)$ be the received sequence. If the adversary is in the eavesdropping state, i.e., $\phi(i)=0$, the channel decoder
 draws $g_z(i)$ from $G_z(i)$ and a noise sequence $s_j^N(i)$ from $S_j^N(i)$ to obtain
 $$\hat{y}^N(i) =y^{N}(i)+g_z(i)s_j^N(i).$$

The channel decoder collects $y^N(i)$'s that correspond to the same bit group and apply MRC to these observations as explained in the proof sketch. Then, the channel decoder  employs joint typicality decoding as in the no feedback case (mentioned in the Appendix~\ref{section:nocsi}).

\emph{Secrecy Analysis:}
For the secrecy analysis,  let's define $\hat{Z}^N(i) =X^{N}(i) g_e(i)+S_e^{N}(i)$, $1 \leq \forall i \leq M$. 
The equivocation analysis averaged over codebooks is as follows: \allowdisplaybreaks
\begin{align}
&\mathbb E_\mathcal{C}[R_e(\mathcal{C})]=\frac{1}{MN}H(W|\{Z^{N}(i)\}_{i:\phi(i)=0}, g^M, 
\mathcal{C}) \\
&\stackrel{(a)}{\geq}\frac{1}{MN} H(W|\hat{Z}^{NM}, g^M,\mathcal{C}) \nonumber \\
&= \frac{1}{MN}H\left(W, \{X^{N}(i)\}_{i:i\in A}|\hat{Z}^{NM},g^M, \mathcal{C}\right)\nonumber\\
&\qquad-\frac{1}{MN}H\left( \{X^{N}(i)\}_{i:i\in A}|\hat{Z}^{NM}, W, g^M,\mathcal{C}\right)\\
&\geq \frac{1}{MN}H\left(\{X^{N}(i)\}_{i:i\in A}|\mathcal{C}\right)\nonumber\\
&\qquad-\frac{1}{MN}I\left(\{X^{N}(i)\}_{i:i\in A};\hat{Z}^{NM}|g^M, \mathcal{C}\right)\nonumber\\
&\qquad\qquad-\frac{1}{MN}H\left( \{X^{N}(i)\}_{i:i\in A}|\hat{Z}^{NM}, W, g^M,\mathcal{C}\right) \nonumber\\
&=\frac{1}{MN}\sum_{i\in A}  H\left(X^N(i)|\hat{Z}^{NM},\mathcal{C},g^M\right)\nonumber\\
&\qquad -\frac{1}{MN}H\left( \{X^{N}(i)\}_{i:i\in A}|\hat{Z}^{NM}, W, g^M,\mathcal{C}\right) \nonumber\\
&=\frac{1}{MN}\sum_{i\in A} \left[H(X^N(i))- I\left(X^N(i);\hat{Z}^{NM}|\mathcal{C},g^M\right)\right]^{+}\nonumber\\
&\qquad-\frac{1}{MN}H\left( \{X^{N}(i)\}_{i:i\in A}|\hat{Z}^{NM}, W, g^M,\mathcal{C}\right)\nonumber\\
&\hspace{-0.5cm}\stackrel{(b)}{=}\frac{1}{MN} \sum_{i\in A}\left[NR-I\left(X^N(i); \hat{Z}^N(i-r(i)+1:i)|\mathcal{C},g^M\right)\right]^{+} \nonumber\\
&\;-\frac{1}{MN}H\left( \{X^{N}(i)\}_{i:i\in A}|\hat{Z}^{NM}, W, g^M,\mathcal{C}\right)\label{l7}\\
&\hspace{-0.5cm}\geq\frac{1}{MN} \sum_{i\in A}\left[NR-I\left(X^N(i),\mathcal{C}; \hat{Z}^N(i-r(i)+1:i)|g^M\right)\right]^{+}\nonumber\\
&\qquad-\frac{1}{MN}H\left( \{X^{N}(i)\}_{i:i\in A}|\hat{Z}^{NM}, W, g^M,\mathcal{C}\right)\nonumber\\
&\stackrel{(c)}{=}\frac{1}{MN} \sum_{i\in A}\left[NR-I\left(X^N; \hat{Z}^N(i-r(i)+1:i)|g^M\right)\right]^{+}\nonumber\\
&\;-\frac{1}{MN}H\left( \{X^{N}(i)\}_{i:i\in A}|\hat{Z}^{NM}, W, g^M,\mathcal{C}\right)\label{l8}\\
&\stackrel{(d)}{=} \frac{1}{M}\sum_{i\in A}\left[R-I\left(X; \hat{Z}_{(i-r(i)+1)\dots, \hat{Z}_i }|g^M\right)\right]^{+}\nonumber\\
&\;-\frac{1}{MN}H\left( \{X^{N}(i)\}_{i:i\in A}|\hat{Z}^{NM}, W, g^M,\mathcal{C}\right) \label{l9}\\
&=\frac{1}{M}\sum_{i\in A} \left[R- \log\left(1+P_t\sum_{j=1}^{r(i)} h_{e}\left(i-j+1\right)\right) \right]^{+}\nonumber\\
&-\frac{1}{MN}H\left( \{X^{N}(i)\}_{i:i\in A}|\hat{Z}^{NM}, W, g^M,\mathcal{C}\right)\label{l10}\\
&\stackrel{(e)}{\geq} C_s^{-\text{1-bit}} \nonumber\\
&\quad-\frac{1}{MN}H\left( \{X^{N}(i)\}_{i:i\in A}|\hat{Z}^{NM}, W, g^M,\mathcal{C}\right)-\epsilon\label{l11}\\
&\geq  C_s^{-\text{1-bit}}-2\epsilon \label{l12}
\end{align}
for any $\epsilon > 0$ and for sufficiently large $M$, where $r(i)$ is the required number of transmissions for the bit group that is successfully decoded on  $i$-th block and $A$ is the set of blocks on which decoding occurs successfully, i.e.,  $A=\left\{i:\log\left(1+\sum_{j=1}^{r(i)-1}\frac{P_th_m(i-j)}{1+P_jh_z(i-j)}\right)<R\leq\log\left(1+\sum_{j=1}^{r(i)}\frac{P_th_m(i-j+1)}{1+P_jh_z(i-j+1)}\right) \text{ and } 1\leq i \leq M\right\}$. Here, $(a)$ follows from the fact that conditioning reduces the entropy. In~\eqref{l7}, $\hat{Z}^N(i-r(i)+1:i)=\left[\hat{Z}^N(i) \dots, \hat{Z}^N(i-r(i)+1)\right]$ is the vector of the observed signals at the adversary that corresponds to successfully received codeword $X^N(i)$. Here, $(b)$ follows from the fact that $X^N(i)$ and $\{Z^N(j)\}_{j\notin (i-r(i)+1, \ldots, i)}$ are independent. In \eqref{l8},  $(c)$ follows from the fact that $\mathcal{C} \to  X^N(i) \to \hat{Z}^N(i),\ldots \hat{Z}^N(i-k+1)$ forms Markov chain. Here, $X^N(i)$ is not conditioned to codebook $\mathcal{C}$, hence $X^N(i)=X^N \sim \mathcal{C}\mathcal{N}(0, P_tI_{N\times N})$. In~\eqref{l9}, 
\begin{equation}
\hat{Z}_k \triangleq X+N_k, \;k\in\{i-r(i)+1,\dots, i\}
\end{equation}
where $N_k$'s are i.i.d and $X$ and $N_k$ are distributed with $\mathcal{C}\mathcal{N}(0,P_t)$ and $\mathcal{C}\mathcal{N}(0,1)$, respectively.
In~\eqref{l9}, $(d)$ follows from the fact that 
\begin{align}
&p_{X^N, \hat{Z}^N\left(i-k+1:i\right)}\left(x^N, z^N(i-k+1:i)\right)=\nonumber\\
&\hspace{-0.35cm}\prod_{j=1}^Np_X(x_j)p_{\hat{Z}_{(i-k+1:i)}}\left(z_j\left(i-k+1:i\right)|x_j,g\left(i-k+1:i\right)\right)\nonumber
\end{align}
where $z_j(i)$ denotes $j$-th element of $i$-th block.  In~\eqref{l10}, $(e)$ follows from the renewal reward theorem. We can show that the second term in \eqref{l11} goes to zero as $M\to \infty$  with the list decoding argument used in the proof of Theorem 2 of \cite{lai2008}. This concludes the proof. \qed

We now give the proof for Corollary~\ref{1bitrate}. Since the proof is similar to the achievability proof of Theorem~\ref{1bitrate_accum}, we only present the differences in codebook generation, encoding, decoding, and secrecy analysis steps.  In the codebook generation, $R_m$ is selected as $R_m = Rp-\epsilon$, where $p$ is defined in Theorem~\ref{1bitrate_accum}. Note that $p=1/\mathbb E[T^{*}]$.

In the encoding step, we select $M$ such that 
$\lvert\frac{1}{MN} R^{**}_w(M)- \frac{R}{\mathbb E[K]}\rvert \leq \epsilon$ with probability 1. Here, $R^{**}_w(M)$ is the accumulated reward at the receiver up to $M$-th block for the renewal process whose inter-renewal time is distributed with $T^{*}$ and  whose rewards at each renewal are $NR$ bits.

In the decoding step, as opposed to the MRC approach, the receiver discards the received sequence, $y^N(i)$ if event $S^c(i)=\left\{\log\left(1+\frac{P_th_m(i)}{1+P_jh_z(i)}\right) <R\right\}$ occurs. Consequently, the transmitter sends back a NAK signal. The receiver successfully decodes a bit group on $i$-th block if event $S(i)$ occurs.

The secrecy analysis is same with the secrecy analysis in Theorem~\ref{1bitrate_accum}. \qed

We now provide the proof of the upper bound in Theorem~\ref{1bitrate}. Instead of an arbitrary adversary strategy, we assume the adversary strategy on a block, $\phi(i)$ is a deterministic function of the instantaneous channel gains on the block, i.e., $\phi(i)=f\left(g_m(i),g_e(i), g_z(i)\right)$. Since we constrain the adversary strategy, the secrecy capacity upper bound for this case is also the upper bound of the secrecy capacity of the original case in which the adversary strategy arbitrarily changes from one block to the next.

Suppose that $R_s$ is an achievable secrecy rate. From definition~\eqref{cond2}, Fano's inequality and the analysis (\ref{llll1}-\ref{sketch_def2}), we have
\begin{align}
&\frac{1}{NM} H(W| Z^{NM},K^{MN}, G_{m}^M,G_{e}^M, G_{z}^M, \Phi^M)\geq R_s- \delta_{NM}\\
&\frac{1}{NM} H(W| Y^{NM}, K^{MN}, G_{m}^M,G_{e}^M, G_{z}^M, \Phi^M) \leq \epsilon_{NM} 
\end{align}
for any deterministic function, $f: \mathbb R\times \mathbb R\times \mathbb R \to [0,1]$. Here, $\Phi(i) = f\left(H_m(i), H_e(i), H_z(i)\right)$ and $\epsilon_{NM}$ and $\delta_{NM}$ go to zero as $N\to \infty$ and $M\to \infty$.  The upper bound follows with following steps.
\allowdisplaybreaks
\begin{align}
&R_s \leq \frac{1}{MN}\min_{f}  H\left(W| Z^{NM},K^{MN},G_{m}^M,G_{e}^M, G_{z}^M, \Phi^M\right)\nonumber\\
&\quad-H\left(W| Y^{NM}, K^{MN},G_{m}^M,G_{e}^M, G_{z}^M, \Phi^M\right) +\gamma_{NM}\\
&\leq\frac{1}{MN}\min_f I\left(W; Y^{NM}|Z^{NM},K^{MN},G_{m}^M,G_{e}^M, G_{z}^M, \Phi^M\right)\nonumber\\
&\qquad\qquad\qquad\qquad\qquad\qquad\qquad\qquad\qquad +\gamma_{NM} \nonumber\\
&\leq I\left(W; Y^{NM}, G_m^{MN},G_z^{MN}|Z^{NM},K^{MN},G_e^{MN}, \Phi^M\right)+\gamma_{NM}\label{mut_term}
\end{align}
where $\gamma_{NM} = \delta_{NM} + \epsilon_{NM}$ and $\gamma_{NM} \to 0$ as $N\to \infty$ and $M\to\infty$. By using the following lemmas, we can reduce the mutual information term in~\eqref{mut_term} to a simplier form. Since Lemma~\ref{ashish_lem1} and Lemma~\ref{ashish_lem2} are similar to Lemma 1 and Lemma 2 of \cite{ashish2013}, respectively, we omit the proofs.
\begin{lemma}\label{ashish_lem1}
For each block $i\in\{1,\dots,M\}$, we have that
\begin{align}
&I\left(W; Y^{Ni}, G_m^M,G_z^M|Z^{Ni},K^{Ni},G_e^M,\Phi^M\right)\leq \nonumber\\
&\quad\;\; I\left(W; Y^{Ni}, G_m^M,G_z^M|Z^{Ni},K^{N(i-1)},G_e^M,\Phi^M\right)
\end{align}\qed
\end{lemma}
\begin{lemma}\label{ashish_lem2}
For each block $i\in\{1,\dots,M\}$, we have that
\begin{align}
&I\left(W; Y^{Ni}, G_m^M,G_z^M|Z^{Ni},K^{N(i-1)},G_e^M,\Phi^M\right)\leq \nonumber\\
&I\left(W; Y^{N(i-1)}, G_m^M,G_z^M|Z^{N(i-1)},K^{N(i-1)},G_e^M,\Phi^M\right)\nonumber\\
&+I\left(X^N(i);Y^N(i)|Z^N(i), G_m(i),G_e(i),G_z(i), \Phi(i)\right)
\end{align}\qed
\end{lemma}
As in~\cite{ashish2013}, by successively applying Lemma 1 and Lemma 2, we can show the following inequality.
\begin{align}
&I\left(W; Y^{NM}, G_m^{MN},G_z^{MN}|Z^{NM},K^{MN},G_e^{MN}, \Phi^M\right)\leq \nonumber \\
&\sum_{i=1}^M I\left(X^N(i);Y^N(i)|Z^N(i), G_m(i),G_e(i),G_z(i), \Phi(i)\right). \nonumber
\end{align}
Hence, we have 
\begin{align}
&R_s-\gamma_{NM}\\
&\leq\frac{1}{MN}\min_f\sum_{i=1}^M I\left(X^N(i);Y^N(i)|Z^N(i), G(i), \Phi(i)\right)\nonumber\\
&\leq\frac{1}{MN}\min_f\sum_{i=1}^M\sum_{j=1}^N I\left(X(i,j);Y(i,j)|Z(i,j), G(i), \Phi(i)\right)\nonumber\\
&\stackrel{(a)}{\leq} \min_f \frac{1}{MN} \sum_{i=1}^M\sum_{j=1}^N \left(\mathbb E\left[\log\left(1+\frac{P_{t_{ij}}H_m(i)}{1+P_jH_z(i)}\right)\bigg \vert f(G(i))=1\right] \mathbb P\left(f(G(i))=1\right)\;\right.\nonumber\\
&\qquad\qquad\qquad\qquad\qquad \qquad +\left. \mathbb E\left[\log\left(1+\frac{P_{t_{ij}}H_m(i)}{1+P_{t_{ij}}H_e(i)}\right)\bigg\vert f(G(i))=0\right] \mathbb P(f(G(i))=0)\right)\\
&\stackrel{(b)}{\leq} \min_f  \bigg{(}\mathbb E\left[\log\left(1+\frac{\frac{1}{MN} \sum_{i=1}^M\sum_{j=1}^NP_{t_{ij}}H_m}{1+P_jH_z}\right)\bigg \vert f(G)=1\right] \mathbb P\left(f(G)=1\right)\;\nonumber\\
&\qquad\qquad\qquad\qquad \qquad + \mathbb E\left[\log\left(1+\frac{\frac{1}{MN}\sum_{i=1}^M\sum_{j=1}^N P_{t_{ij}}H_m}{1+\frac{1}{MN} \sum_{i=1}^M\sum_{j=1}^NP_{t_{ij}}H_e}\right)\bigg\vert f(G)=0\right] \mathbb P(f(G)=0)\bigg{)}\label{jensen_2}\\
&\stackrel{(c)}{\leq} \min_f  \bigg{(}\mathbb E\left[\log\left(1+\frac{P_tH_m}{1+P_jH_z}\right)\bigg \vert f(G)=1\right] \mathbb P\left(f(G)=1\right)\;\nonumber\\
&\qquad\qquad\qquad\qquad \qquad + \mathbb E\left[\log\left(1+\frac{P_tH_m}{1+ P_tH_e}\right)\bigg\vert f(G)=0\right] \mathbb P(f(G)=0)\bigg{)}\label{inc_fun}\\
&= \min_f\mathbb E\left[\log\left(1+\frac{P_tH_m}{1+P_jH_zf(G)+P_tH_e (1-f(G))}\right)\right]\label{max_bef}\\
&\stackrel{(d)}{=}\mathbb E\left[\log\left(1+\frac{P_tH_m}{1+\max\left(P_tH_e,P_jH_z\right)}\right)\right], \label{max_pro}
\end{align}
where the notation $(i,j)$ indicates the $j$-th channel use of $i$-th block, $G(i)=\left[G_m(i),G_e(i),G_z(i)\right]$, $G=\left[G_m,G_e,G_z\right]$, and $H=\left[H_m,H_e,H_z\right]$. The power constraint in~\eqref{pw_cst} implies that  $\frac{1}{NM}\sum_{i=1}^{M}\sum_{j=1}^N \mathbb E\left[|X(i,j)|^2\right]\leq P_t$, where the expectation is taken over $W$ and $K^{(i-1)N}$. Also, note that 
$G(i)=\left[G_m(i), G_e(i),G_z(i)\right]$ and $X(i,j)$ are independent random variables. Define $P_{{t}_{ij}}\triangleq\mathbb E\left[|X(i,j)|^2\right]= \mathbb E\left[|X(i,j)|^2|G(i)=g(i)\right]$. Then, $(a)$ follows from the fact that Gaussian distribution maximizes the conditional mutual information~\cite{ashish2013} for both values of $\Phi(i)$. In~\eqref{jensen_2}, $(b)$ follows from   Jensen's inequality and from the fact that $\log\left(1+P_{t_{ij}}x\right)$ and $\log\left(1+\frac{P_{t_{ij}}x}{1+P_{t_{ij}}y}\right)$ are concave functions of $P_{t_{ij}}$ for any $x\geq 0$ and $y\geq0$. In~\eqref{inc_fun}, $(c)$ follows from the fact that  $\left(\log(1+Px\right)$ and $\log\left(1+\frac{Px}{1+Py}\right)$ are non-decreasing functions in $P$ for any $x\geq 0$ and $y\geq0$. In~\eqref{max_pro}, $(d)$ follows from the fact that $f(G)= I_{P_jH_z\geq P_tH_e}$ minimizes the expectation in \eqref{max_bef}, where $I_{x\geq a} =1$ if $x\geq a$; otherwise, $I_{x\geq a} =0$.
\qed
\section{Proof of Theorem~\ref{t:delaynofeedback}}\label{app:delay_nofb}
Fix $\gamma \in [0,1]$, $\bar{\gamma}=1-\gamma$, $\epsilon>0$. Each consecutive $M_1$ blocks is called a superblock. Suppose that communication lasts $M=M_1M_2$ blocks. Let us denote $x^{NM_1}(j)$, $y^{NM_1}(j)$, and $z^{NM_1}(j)$ as the transmitted signal, the received signal at the receiver, and the received signal at the adversary in superblock $j$, respectively. Denote  $x^{\gamma N}(j,i)$ and $x^{\bar{\gamma} N}(j,i)$ as the transmitted signals in the first $\gamma N$ channel uses and in the  next $\bar{\gamma}N$ channel uses of $i$-th block of $j$-th superblock, respectively. Signals $y^{\gamma N}(j,i)$, $y^{\bar{\gamma} N}(j,i)$, $z^{\gamma N}(j,i)$ and $z^{\bar{\gamma} N}(j,i)$ are defined in a similar way. Let $w(j,i)$ be the message to be transmitted in $i$-th block of $j$-th superblock. Finally, let $x^{\gamma NM_1}(j)\triangleq[x^{\gamma N}(j,1),\dots,x^{\gamma N}(j,M_1)]$, and $y^{\gamma NM_1}(j),z^{\gamma NM_1}(j),x^{\bar{\gamma} NM_1}(j),y^{\bar{\gamma} NM_1}(j), \text{ and } x^{\bar{\gamma} NM_1}(j)$ are defined in a similar way. Through this appendix, $(j,i)$ indicates $i$-th block of $j$-th superblock.
\begin{figure}[t]
   \centering
   \includegraphics[width=0.5\textwidth]{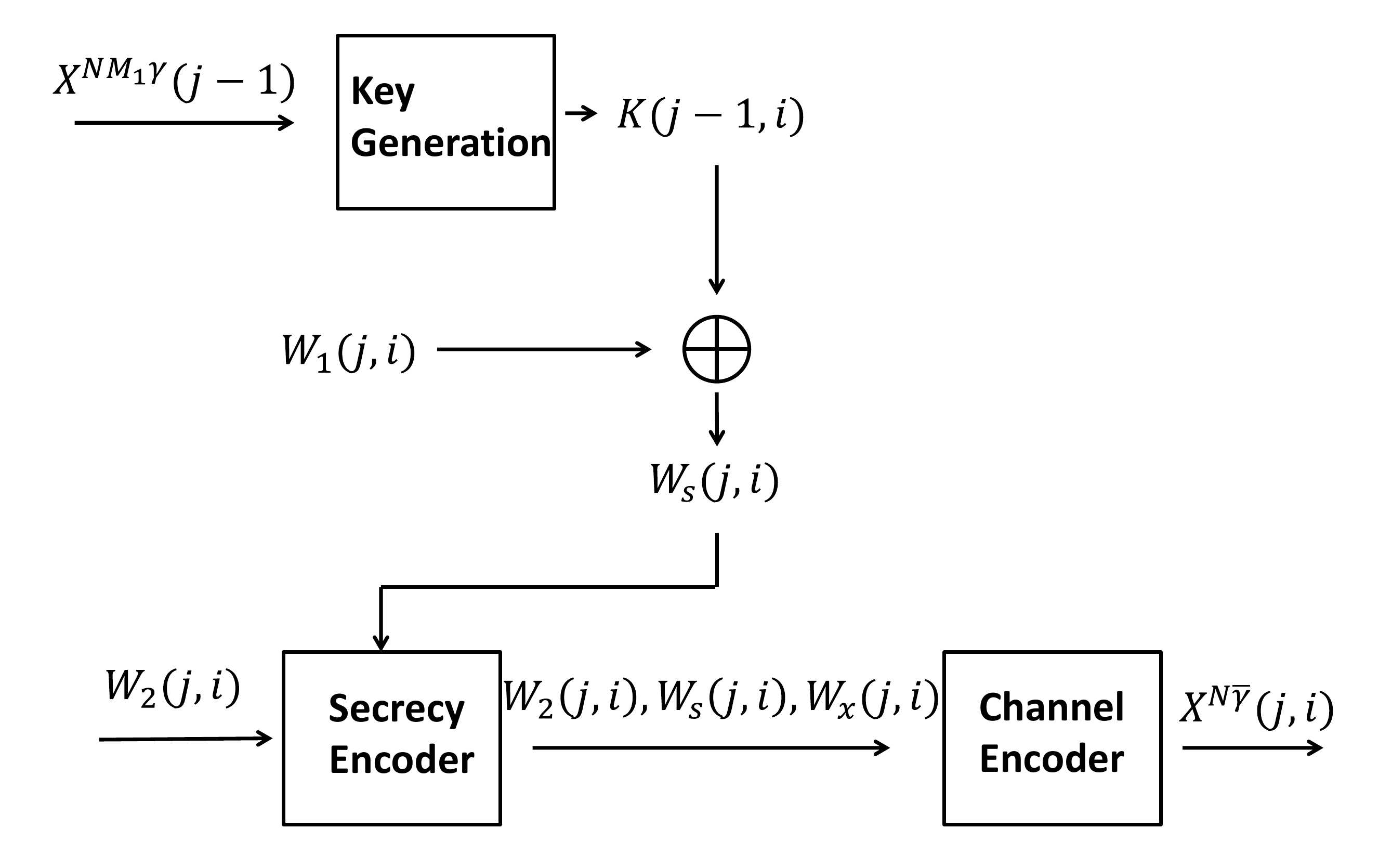}
   \caption{Encoder structure.}
   \label{fig:delay1}
 \end{figure}
\begin{figure}[t]
   \centering
   \includegraphics[width=0.5\textwidth]{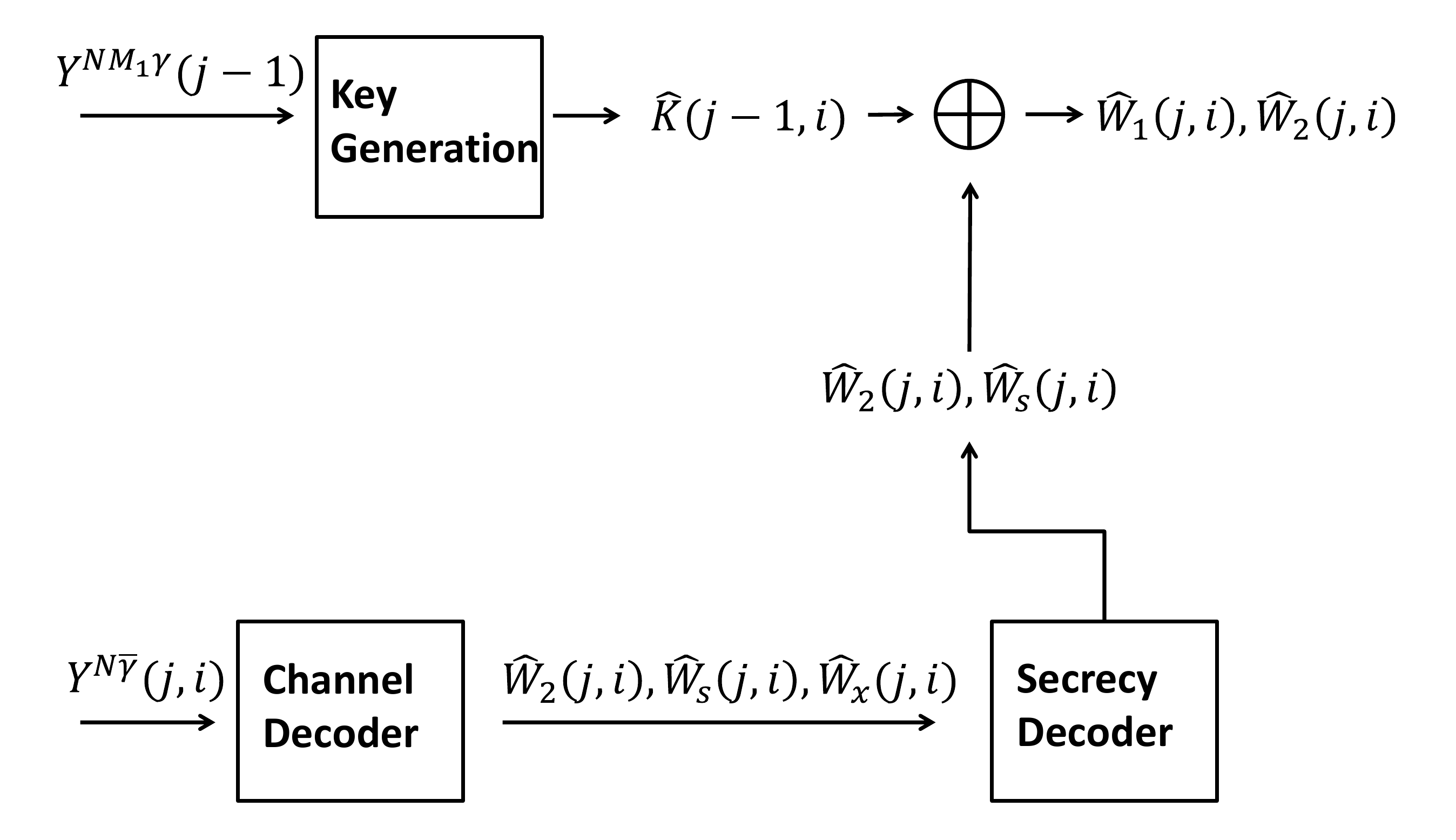}
   \caption{Decoder structure.}
   \label{fig:delay2}
 \end{figure}

Encoding and decoding strategies are summarized in Figure~\ref{fig:delay1} and Figure~\ref{fig:delay2}. We begin with key generation. Let $R_{r0}>0$.  At the beginning of superblock $j$, the transmitter picks key $k(j)$ from random variable $K(j)$ which is uniformly distributed in $\{1,\ldots,2^{NR_{r0}}\}$. By using the encoding strategy in the proof of Theorem 1, the transmitter maps $k(j)$ to codeword $x^{\gamma NM_1}(j)$. This process is repeated for every superblock $j\geq 1$. Next lemma provides a lower bound to achievable key rates. 
\begin{lemma} \label{mod_nofb}
 For any $\epsilon > 0$, there exit $N'>0$, $M'_1 > 0$ and a sequence of length 
$\gamma N{M_1}$ channel codes $\left(\gamma NM_1, 2^{\gamma R_{r0}NM_1}\right)$ for which the following are satisfied 
under any strategy of the adversary, $\phi^{M_1}(j)$:
\begin{align}
&\Pr\left(K(j)\neq\hat{K}(j)\right) < \epsilon /3   \label{appdelay:keylemma1}\\
&\frac{1}{NM_1}H\left(K(j)|\{Z^{\gamma NM_1}(j)\}, g^{M_1}(j),\phi^{M_1}(j)\right) > R_{r0} - \epsilon/2 \label{appdelay:keylemma2} 
\end{align}
for any superblock $j\in \{1,2,\dots, M_2\}$,  for any $N\geq N'$, and for any $M_1\geq M_1'$ where $R_{r0}\leq \gamma C_s^{-}$.\qed
\end{lemma}
The proof follows from Theorem 1. Now, we describe the transmission of  delay limited message $w(j,i)$\footnote{Due to Definition~\ref{secrecyoutageconstraint}, 
we skip the message transmission at first $M_1$ blocks, and declare \emph{secrecy outage}.}, illustrated in Figure~\ref{fig:delay1}. Let $R_s \geq R_{r0}$ and $\tilde{R}_s \geq R_{s}$.
Message $w(j,i)$ of size $NR_s$ bits
is divided\footnote{Note that in this process, the messages are converted to binary form.} to two messages $w_1(j,i)$ and $w_2(j,i)$, of size $NR_{r0}$ and $N(R_s-R_{r0})$, respectively.
We also divide key $k(j-1)$, generated in previous superblock $j-1$, into $M_1$ equivalent size chunks such that
$k(j-1) = \left[k(j-1,1)\ldots,k(j-1,M_1)\right]$, where $k(j-1,i)$ is of size $N R_{r0}$ bits.

Let $w_s(j,i) = w_1(j,i) \oplus k(j-1,i)$. Suppose $w_x(j,i)$ is picked from random variable $W_x(j,i)$ which is uniformly distributed on sample space $\{1,\dots, 2^{N-\tilde{R}_s-R_s-\epsilon}\}$ and independent from $W(j,i)$. We generate a Gaussian codebook consisting of $2^{N(\tilde{R}_s-\epsilon)}$ codewords each of which are independently drawn from $\prod_{k=1}^{\bar{\gamma}N} p_X(x_k)$. Here, $p_X(x)$ is the probability density function of complex Gaussian random variable with zero mean and variance $P_t$. To transmit $w(j,i)=\left(w_1(j,i), w_2(j,i)\right)$, the codeword indexed by $(w_1(j,i), w_s(j,i), w_x(j,i))$ is transmitted.

\emph{Error and Equivocation Analysis}:
\begin{lemma}\label{mod_cap}
For any $\epsilon > 0$, there exit $N''>0$  and a sequence of length 
$\bar{\gamma} N$ channel codes $(\bar{\gamma}N, 2^{\bar{\gamma} \tilde{R}_sN})$ for which the following are satisfied 
\begin{align}
&\Pr\big{(}(W_2(j,i),W_s(j,i),W_x(j,i))\neq (\hat{W}_2(j,i),\hat{W}_s(j,i),\hat{W}_x(j,i)) \big{)}  < \epsilon/3
\end{align}
for any $j\in \{1,2,\dots M_2\}$, for any $i\in\{1,2,\dots M_1\}$ and for any $N \geq N''$ when the channel conditions satisfy 
\begin{align}
\bar{\gamma}\log\bigg{(}1+\frac{P_th_m(j,i)}{1+P_jh_z(j,i)}\bigg{)}\geq \tilde{R}_s \label{app:condition1}.
\end{align}\qed
\end{lemma}
The proof follows from standard arguments, and is omitted. 
Assume for the error and equivocation analysis that $N$ and $M_1$ are chosen such that
$N=\max(N',N'',N''')$, and $M_1=M_1'$, where $N'''$ will be defined later.
Then, error probability is bounded as
\begin{align}
&\Pr(E(j,i))  \triangleq  \Pr\big{(}(W_1(j,i),W_2(j,i)) \neq (\hat{W}_1(j,i),\hat{W}_2(j,i))\big{)} \nonumber \\
		  & \leq \Pr\left(\left(W_2(j,i)\neq \hat{W}_2(j,i)\right) \bigcup \left(W_1(j,i)\neq \hat{W}_1(j,i)\right)\right) \nonumber\\
&\leq \frac{\epsilon}{3}+\Pr\left(W_1(j,i)\neq \hat{W}_1(j,i)\right)\label{er1}\\
&\leq \frac{\epsilon}{3}+\Pr\left(W_s(j,i)\neq \hat{W}_s(j,i) \bigcup K(j,i)\neq \hat{K}(j,i)\right)\label{er2}\\
&\leq \epsilon, \label{er3}
\end{align}
where \eqref{er1} follows from Lemma~\ref{mod_cap}, \eqref{er2} follows from the fact that $W_1(j,i)=W_s(j,i)\oplus K(j,i)$ and \eqref{er3} follows from Lemma~\ref{mod_nofb} and Lemma~\ref{mod_cap}.

For the secrecy analysis,  let's define $\hat{Z}^N(j,i) =X^{N}(j,i) g_e(j,i)+S_e^{N}(j,i)$, $1 \leq \forall j \leq M_1$, $1 \leq \forall i \leq M_2$.  
Equivocation analysis averaged over codebooks is as follows. Note that all the equivocation terms below are conditioned on the channel gains $g^M$, and we omit them for the sake of simplicity.
 \begin{align}
&H( W_1(j,i),W_2(j,i)|Z^{NM},W^{M}\backslash W(j,i),\mathcal{C})  \nonumber\\
& \geq H( W_1(j,i),W_2(j,i)|\hat{Z}^{NM},W^{M}\backslash W(j,i),\mathcal{C})  \\
&\;= H(W_2(j,i)|\hat{Z}^{NM},W^{M}\backslash W(j,i),\mathcal{C})\nonumber \\
&\;\;\; + H(W_1(j,i)|\hat{Z}^{NM},W^{M}\backslash W(j,i),W_2(j,i),\mathcal{C})   \label{appdelay:eq0}  
\end{align}
We now bound the first term in \eqref{appdelay:eq0}.
\begin{align}
&H(W_2(j,i)|\hat{Z}^{NM},W^{M}\backslash W(j,i),\mathcal{C})  \nonumber\\
		&=H(W_2(j,i)) -I(W_2(j,i);\hat{Z}^{NM},W^{M}\backslash W(j,i)|\mathcal{C})\label{del_part1_0}\\
&=H(W_2(j,i)) -I(W_2(j,i);\hat{Z}^{N\gamma}(j-1),\hat{Z}^{NM_1\bar{\gamma}}(j), W^{M_1}(j)\backslash W(j,i) |\mathcal{C})\label{del_part1_1}\\
&= H(W_2(j,i)) -I(W_2(j,i);\hat{Z}^{N\bar{\gamma}}(j,i)|\mathcal{C})\nonumber\\
&-I(W_2(j,i);\hat{Z}^{NM_1\gamma}(j-1),\hat{Z}^{NM_1\bar{\gamma}}(j)\backslash \hat{Z}^{N\bar{\gamma}}(j,i), W^{M_1}(j)\backslash W(j,i) |\hat{Z}^{N\bar{\gamma}}(j,i),\mathcal{C})\\
&\geq H(W_2(j,i)) -I(W_2(j,i);\hat{Z}^{N\bar{\gamma}}(j,i)|\mathcal{C})\nonumber\\
&-I(W_2(j,i);K(j-1,i),\hat{Z}^{NM_1\gamma}(j-1),\hat{Z}^{NM_1\bar{\gamma}}(j)\backslash \hat{Z}^{N\bar{\gamma}}(j,i), W^{M_1}(j)\backslash W(j,i) |\hat{Z}^{N\bar{\gamma}}(j,i),\mathcal{C})\\
&= H(W_2(j,i)) -I(W_2(j,i);\hat{Z}^{N\bar{\gamma}}(j,i)|\mathcal{C})\nonumber\\
&-I(W_2(j,i);K(j-1,i),\hat{Z}^{NM_1\gamma}(j-1)|\hat{Z}^{NM_1\bar{\gamma}}(j),W^{M_1}(j)\backslash W(j,i),\mathcal{C})\\
&= H(W_2(j,i)) -I(W_2(j,i);\hat{Z}^{N\bar{\gamma}}(j,i)|\mathcal{C})\nonumber\\
&-I(W_2(j,i);K(j-1,i)|\hat{Z}^{NM_1\bar{\gamma}}(j),W^{M_1}(j)\backslash W(j,i),\mathcal{C})\nonumber\\
&\;\;-I(W_2(j,i);\hat{Z}^{NM_1\gamma}(j-1)|K(j-1,i),\hat{Z}^{NM_1\bar{\gamma}}(j),W^{M_1}(j)\backslash W(j,i),\mathcal{C})\\
&= H(W_2(j,i)) -I(W_2(j,i);\hat{Z}^{N\bar{\gamma}}(j,i)|\mathcal{C})\nonumber\\
&\;\;-I(W_2(j,i);\hat{Z}^{NM_1\gamma}(j-1)|K(j-1,i),\hat{Z}^{NM_1\bar{\gamma}}(j),W^{M_1}(j)\backslash W(j,i),\mathcal{C}) \label{del_part1_2}\\
&= H(W_2(j,i)) -I(W_2(j,i);\hat{Z}^{N\bar{\gamma}}(j,i)|\mathcal{C})\label{del_part1_3}\\
& = H(W_2(j,i)|\hat{Z}^{N\bar{\gamma}}(j,i),\mathcal{C})\nonumber\\
&\geq R_s-R_{r0} -N\epsilon/2, \label{del_part1_4}
\end{align}				
where \eqref{del_part1_1} follows from the fact that $\hat{Z}^{NM}\backslash \left(\hat{Z}^{NM_1\gamma}(j-1),\hat{Z}^{NM_1\bar{\gamma}}(j)\right), W^{M}\backslash W^{M_1}(j)$ are independent from the rest of the random variables in~\eqref{del_part1_0} for every codebook. \eqref{del_part1_2} follows from the fact that $K(j-1,i)$ and $\left(W_2(j,i),\hat{Z}^{NM_1\bar{\gamma}}(j),W^{M_1}(j)\backslash W(j,i),\mathcal{C}\right)$ are independent due to the fact that $K(j-1,i) \to W_1(j,i)\oplus K(j-1,i), \mathcal{C}\to W_2(j,i),\hat{Z}^{NM_1\bar{\gamma}}(j),W^{M_1}(j)\backslash W(j,i),\mathcal{C}$ forms Markov chain, and $K(j-1,i)$ and $\left(W_1(j,i)\oplus K(j-1,i), \mathcal{C}\right)$ are independent. \eqref{del_part1_3} follows from the fact $\hat{Z}^{NM_1\gamma}(j-1)\to K(j-1,i),\mathcal{C} \to W_2(j,i),\hat{Z}^{NM_1\bar{\gamma}}(j),W^{M_1}\backslash W(j,i)$ forms Markov chain. Following the same steps in the equivocation analysis in Theorem~\ref{thm:nocsi}, we can show that \eqref{del_part1_4} is satisfied for any $N \geq N^{'''}$ if $\tilde{R}_s- (R_s-R_{r0})\geq \log(1+Ph_e(j,i))$.  Next, we bound the second term in~\eqref{appdelay:eq0}.
\begin{align}
&H( W_1(j,i)|\hat{Z}^{NM},W^{M}\backslash W(j,i),W_2(j,i),\mathcal{C}) \label{del_part2_00}\nonumber\\
&=H(W_1(j,i)|\hat{Z}^{NM_1\gamma}(j-1),\hat{Z}^{NM_1\bar{\gamma}}(j), W^{M_1}(j)\backslash W(j,i),W_2(i,j),\mathcal{C})\\
		&\geq H(W_1(j,i)|\hat{Z}^{NM_1\gamma}(j-1),\hat{Z}^{NM_1\bar{\gamma}}(j), A)\label{del_part2_0} \\
		&=   H(K(j-1,i)|\hat{Z}^{NM_1\gamma}(j-1),\hat{Z}^{NM_1\bar{\gamma}}(j), A)\label{del_part2_1} \\
		&= H(K(j-1,i)) -I(K(j-1,i);\hat{Z}^{NM_1\gamma}(j-1),\hat{Z}^{NM_1\bar{\gamma}}(j), A) \\
		&=  H(K(j-1,i))-I(K(j-1,i);\hat{Z}^{NM_1\gamma}(j-1),\hat{Z}^{NM_1\bar{\gamma}}(j)|A)\label{del_part2_2}\\
		&=  H(K(j-1,i))-I(K(j-1,i);\hat{Z}^{NM_1\gamma}(j-1)|A)\nonumber\\
       &-I(K(j-1,i);\hat{Z}^{NM_1\bar{\gamma}}(j)|\hat{Z}^{NM_1\gamma}(j-1),A)\\
		&=  H(K(j-1,i))-I(K(j-1,i);\hat{Z}^{NM_1\gamma}(j-1)|\mathcal{C})\nonumber\\
       &-I(K(j-1,i);\hat{Z}^{NM_1\bar{\gamma}}(j)|\hat{Z}^{NM_1\gamma}(j-1),A)\label{del_part2_rev1}\\
		&=  H(K(j-1,i))-N\epsilon/2\nonumber\\
      &-I(K(j-1,i);\hat{Z}^{NM_1\bar{\gamma}}(j)|\hat{Z}^{NM_1\gamma}(j-1),A)\label{del_part2_3}\\
		&= N R_{r0} - N\epsilon/2 \label{del_part2_4} ,
\end{align}
where $A = \left(W^{M_1}(j)\backslash W(j,i),W_2(j,i),W_s(j,i),\mathcal{C}\right)$. Here, \eqref{del_part2_0} follows from the fact in~\eqref{del_part1_1}, \eqref{del_part2_1} follows from the fact that $W_s(j,i) = W_1(j,i) \oplus K(j-1,i)$, \eqref{del_part2_2} follows from the fact that  $K(j-1,i)$ and $A$ are independent, and \eqref{del_part2_rev1} follows from the fact that  $\left(K(j-1), \hat{Z}^{NM_1\gamma}(j-1), \mathcal{C}\right)$ are independent of $\left(W^{M_1}(j)\backslash W(j,i),W_2(j,i),W_s(j,i)\right)$. From Lemma~\ref{mod_nofb}, we observe that
\eqref{del_part2_3} is satisfied for any $N\geq N^{'}$ and for any $M\geq M_1'$  if $R_{r0} \leq \gamma C_s^{-}$.  \eqref{del_part2_4} follows from the fact that $K(j-1,i) \to\hat{Z}^{NM_1\gamma}(j-1),A \to  \hat{Z}^{NM_1\bar{\gamma}}(j)$ forms Markov chain. Combining \eqref{del_part1_4} and \eqref{del_part2_4}, we can observe 
\begin{align}
H(W_1(i),W_2(i)|Z^{NM},W^{NM}\backslash W(i)) \geq N(R_s-\epsilon),														
\end{align}
if $\tilde{R}_s- (R_s-R_{r0})\leq \log(1+Ph_e(j,i))$ and $R_{r0} \leq \min(\gamma C_s^{-}, R_s)$. 

We can observe that $\alpha$-outage secrecy capacity is lower bounded by $R_s$ if there exists $\left(R_s, \tilde{R}_s, R_{r0},\gamma  \right)$ that satisfy the following conditions:
1) $\Pr\bigg{(} \left\{(1-\gamma) \log\left(1+\frac{P_tH_m}{1+P_jH_z}\right)\geq \tilde{R}_s\right\} \bigcap  
\left\{ \tilde{R}_s-R_s+R_{r0}\geq (1-\gamma)\log(1+P_tH_e) I_{R_s\neq R_{r0}} \right\}\bigg{)}\geq 1-\alpha$, 2) 
$R_{r0}\leq \min\left(R_s,\gamma C_s^{-}\right)$, 3) $R_s\leq \tilde{R}_s$, and 4) $\gamma \in [0,1]$. Notice that the second event in the probability term is equal to $\left\{\left[\tilde{R}_s-(1-\gamma)\log(1+P_tH_e)\right]^{+}\geq R_s-R_{r0} \right\}$. 

Let's define set $A$ containing $\left(R_s, \tilde{R}_s, R_{r0},\gamma  \right)$'s that satisfy these four conditions. The lower bound to $\alpha$ outage secrecy capacity can be written as $C_{s_d}^{-}(\alpha) = \max_{R_s, \tilde{R}_s, R_{r0},\gamma \in A} R_s$. It is easy to observe that if $R_s = C^{-}_{s_d}(\alpha)$, the corresponding $R_{r0}$ has to be equal to $\gamma C_s^{-}$. Then, the lower bound can be written as 
\begin{align}
&C^{-}_{s_d}(\alpha) = \max_{R_s, \tilde{R}_s, R_{r0},\gamma \in A} R_s \\
& \text{subject to } R_{r0} = \gamma C_s^{-} \nonumber\\
\end{align}
which concludes the proof.\qed


\end{document}